\documentclass[twocolumn]{aastex61}

\usepackage[normalem]{ulem}

\definecolor{dark-green}{RGB}{0,100,0}

\newcommand{\numbered}[2]{#2}
\newcommand{\newtext}[1]{#1}
\newcommand{\redout}[1]{}

\newcommand{\group}{\text{group}}
\graphicspath{{./figures/}}

\def\msun{{\rm\,M_\odot}}

\shorttitle{Early Assembly in Galaxy Groups}
\shortauthors{Li and Cen}

\begin{document}

\title{“Assembly Conformity of Structure Growth: Fossil versus Normal Groups of Galaxies}

\correspondingauthor{Zack Li}
\email{zequnl@astro.princeton.edu}

\author[0000-0002-0309-9750]{Zack Li}
\affiliation{Department of Astrophysical Sciences \\
Princeton University \\
Peyton Hall, 4 Ivy Lane \\
Princeton NJ 08544, USA}

\author{Renyue Cen}
\affiliation{Department of Astrophysical Sciences \\
Princeton University \\
Peyton Hall, 4 Ivy Lane \\
Princeton NJ 08544, USA}

\begin{abstract}

Using a semi-analytic method calibrated to the global star formation history and the stellar mass function at $z=0$, we attempt to understand the most stellar deficient galaxy groups.
We argue such groups are a kind of fossil group (FGs) --- in comparison to the normal groups of galaxies, they assemble both halo and stellar mass earlier. We find there is a central galaxy and satellite conformity between these FGs and normal groups: centrals and satellites in the former form earlier and more stellar deficient than their counterparts of the latter. We term this effect ``Assembly Conformity" of dark matter halos. This effect accounts for about 70\% of the difference in stellar content between FGs and normal groups. When split by the peak redshift for the star formation rate of a group, the mass functions of satellite halos on either side of the peak redshift are found to be indistinguishable between FGs and normal groups, indicating a self-similarity of halo assembly with respect to the peak. 
The ``baryonic environmental" effect due to ram-pressure and gas heating accounts for about 30\% of the difference in stellar content. 
While the total stellar mass of FGs is lower than that of normal groups, we predict that the mass of the brightest central galaxy of FGs is, on average, higher than that of normal groups. 
We also predict that in the central galaxies of FGs, there is a negative stellar age gradient from the center outward, where the opposite is expected for those in normal groups.

\end{abstract}

\keywords{}

\section{Introduction} \label{sec:intro}

Fossil groups (FGs) are galaxy systems which have extended X-ray halos and a single bright, central galaxy which dominates the system's optical luminosity. Such groups have been long conjectured to occur when the largest satellites of an old and isolated group have merged into the central galaxy through dynamical friction \citep{barnes1989, donghia2005, dariush2007}. Direct evidence for this picture is still lacking, but there is allure to a connection between group morphology and history, and such relaxed groups would be a useful laboratory for studying galaxy evolution while normalizing for recent mergers.

Although the original motivation was to find early-forming systems, selection criteria for FGs typically rely on the gap in optical magnitude between the first and second brightest galaxies. This selection method stems from the idea that repeated mergers in isolated and relaxed systems should make the central galaxy brighter, while simultaneously depleting the population of bright satellites. \cite{jones2003} established the conventional observational definition for FGs, identifying FGs as extended sources with X-ray luminosity $L_X \geq 10^{42} h_{50}^{-1}$ ergs s$^{-1}$ and an $R$-band magnitude gap $\Delta m_{12} \geq 2.0$ mag between their central galaxy and the second brightest object within half the projected virial radius. 


Despite a conjectured formation scenario which departs from ordinary hierarchical assembly, the only observed properties of FGs which differ significantly from groups of the same halo mass relate to their brightest \newtext{group or} cluster galaxy (BCG). There is a consensus among observers that FG BCGs are \newtext{special}, and in particular brighter than those of typical groups \citep{harrison2012}, dominating the total optical luminosity of the cluster. This result follows from the magnitude-gap selection criteria, but excludes groups which have typical stellar masses but have fewer massive satellites and more low-mass satellites. \cite{bharadwaj2016} find that the X-ray peak is located near the dominant elliptical in all seventeen FGs they studied, further supporting the existence of a connection between FG BCGs and overall group properties. \numbered{6}{\cite{raouf2019} find substantial differences in the color of GAMA brightest group galaxies, using SED-fitting of GAMA groups selected with a magnitude gap and BCG offset, suggesting differences in the central stellar populations of relaxed groups relative to typical groups.}

\redout{Almost all other properties of} \newtext{There is evidence that FGs have stellar population properties consistent with} typical groups. \cite{labarbera2009} study 25 FGs and find similar ages, metallicities, and $\alpha$-enhancements as bright field ellipticals. \cite{eigenthaler2013} measure metallicity gradients of six central galaxies in FGs and find generally negative metallicity gradients similar to those found in cluster ellipticals of similar mass. \cite{zibetti2009} study a sample of six FG and find the cumulative substructure distribution functions (CSDFs) of FGs are consistent with those of typical clusters. \cite{zarattini2016} find galaxy substructures in twelve systems and find a similar fraction of substructure as detected in non-fossil clusters. FGs also appear in both poor and rich environments \citep{pierini2011,adami2012}. Similarly, scaling relations for X-ray luminosity and temperature of FGs match those of non-FGs \citep{khosroshahi2007}. \numbered{6}{Likewise, \cite{trevisan2017} find no trends in stellar population properties like age, abundances, metallicities, and star formation histories of 550 SDSS groups with respect to the magnitude gap.}

There is some debate about whether FGs are less bright in the optical than typical groups at fixed halo mass. Early observations reported halo mass-to-light ratios which were a factor of 2-3 higher than typical groups \citep{vikhlinin1999, yoshioka2004, khosroshahi2007}, but these studies often involved only a handful of systems in their samples. Recent observations with larger sample sizes sometimes corroborate this factor of 2-3 \citep{proctor2011, miller2012, khosroshahi2014}. However, other \redout{recent} studies \citep{voevodkin2010, harrison2012, kundert2015} find no difference between optical and X-ray data, and argue that earlier works had inhomogeneous samples or selection effects.

These disagreements may come from issues with selection criteria, as samples selected through the magnitude gap suffer from both false positives and incompleteness for early-forming systems. Systems which assemble early, but are not completely isolated, can be excluded from the FG sample if they randomly have a moderately bright galaxy in the process of merging. Conversely, systems that randomly have fewer bright satellites can be included in this FG sample, even if they did not form early. \cite{dariush2010} propose an alternate definition involving the gap between the first and fourth brightest galaxies to avoid excluding FGs with a few bright satellites, but find that both definitions fail to find the majority of early-forming systems. They also find that the FG definition is highly transient, with 90\% of FGs becoming non-FGs after $\sim 4$ Gyr.

Possibly also due to these selection issues, there exists evidence for inhomogeneity of baryonic properties within the FG population. \cite{proctor2014} found substantial differences between the age and metallicity gradients of two FGs with similar morphology, luminosity, color, and kinematics, and suggest that FGs do not have homogeneous star formation histories. \cite{bharadwaj2016} investigated the gas dynamics of FGs and found cool cores but a lack of a universal temperature profile, with some of the expected features of relaxed cool-core objects missing in low-temperature examples.

Theoretical work on the origin of FGs generally agree that the appearance of an optical gap is a transient phenomenon \citep{vonbendabeckmann2008, dariush2010, kundert2017}. However, there is some disagreement about whether FGs really form earlier than typical groups, and whether the environment is important in FG formation \citep{donghia2005,diazgimenez2011, cui2011}. \numbered{6}{Studies which suggest a connection between the central galaxy and the mass assembly history \citep{khosroshahi2017, raouf2018} support the idea that the magnitude gap is a good proxy for assembly history. } 


In this paper, we study how the low stellar mass tail of galaxy groups at a given halo mass at $z=0$, which we define as the fossil groups (FGs),
become stellar-mass deficient.
We show, by a systematic dissection of involved processes,
that large scale gravitational growth (i.e., dark matter) physics 
plays the chief role in determining the stellar mass of a group of galaxies.
In particular, we show that 
both the primary progenitor of a group, i.e., the trunk of the merger tree of a group,
and the branches and leaves of the merger tree, display assembly conformity in the sense that
FGs and their building blocks
have, on average, a higher assembly redshift than normal groups and their building blocks 
of the same halo mass at $z=0$.
In contrast, the overall mass distribution of constituent halos and subhalos
in the two types (FGs versus normal) of groups over the entire merger trees are very similar.

We demonstrate that the large scale gravitational effects of assembly conformity cause the main progenitor and massive subhalos of FGs to enter hot mode accretion at a higher redshift than their counterparts of normal groups, shutting out the main progenitor and their massive subhalos of the former from the opportunity to form stars. This process leads to lower stellar content in FGs than in normal groups. Satellite galaxies that are not massive enough to be self-quenching still experience suppressed star formation from these gravitational processes, due to the environmental effects of massive neighbors. This primarily baryonic environmental effect acts in the same way as the gravitational environment effect above, but plays a secondary role in reducing stellar mass in FGs.

From the viewpoint of a superposition of waves of a random gaussian density field,
the physical picture is that the individual density peaks producing the FG primary progenitor and suhalos tend to be higher density
than in normal groups, although the overall peak height at the group scale is similar since we select for a narrow mass range at $z=0$ in our study.
While this finding itself is new and subtle,
it is in some way similar to the large-scale correlation
among neighboring halos at large distances 
noted by \citealt{hearin2016}.

The outline of the paper is as follows.
A description of a physical model for star formation is given in \S \ref{sec:model}.
The results are presented in \S 3, followed
by conclusions in \S 4.



\section{Methods}

\subsection{Star Formation Model} \label{sec:model}

\numbered{7}{Galaxy formation is complicated.
One can try directly simulating the dark matter and gas physics using massive hydrodynamic simulations (i.e. \citealt{vogelsberger2020}), but these still require
parameterizing well-known but poorly constrained processes
such as feedback processes from super-massive black growth (i.e. \citealt{katz1992, springel2005, crain2009, schaye2010, vogelsberger2014, pillepich2018}).
One can sacrifice predictive power for cheaper computational expenses using a more phenomenological approach on dark matter-only simulations,
where one uses a set of parameters to model the physical processes
thought to be relevant for galaxy formation. These are typically referred to as semi-analytic models (SAMs) and process merger trees from N-body simulations into baryonic predictions (i.e. \citealt{kauffmann1993, somerville1999, bower2006, benson2012, henriques2015, raouf2017}).}

We follow in the direction of this semi-analytic method, by implementing 
star formation in halo catalogs extracted from massive N-body simulation, creating a simplified model intended to provide physical understanding.
\numbered{8}{This allows us to easily vary physical effects like the environmental suppression of star formation, which would be much more difficult with the expensive hydrodynamic approach.
By using a simplified model based on the merger trees, we can draw clear conclusions between our modeling and our results on fossil group formation.}
We use the BolshoiP N-body simulations \citep{trujillogomez2011, klypin2011, klypin2016} which has a box size of 0.25 $h^{-1}$ Gpc and 2048$^3$ particles, with particle masses of $1.5 \times 10^8 \, h^{-1} M_{\odot}$ and
a spatial resolution of $1h^{-1}$ kpc comoving. 
The initial conditions are generated with cosmological parameters consistent with \cite{2014PlanckParams}. 
We make a reasonable assumption that the gravitational dynamics of galaxy group size halos
are not significantly impacted by baryonic physics, so merger trees from simulations containing only dark matter are sufficient to model the growth and merger history of halos pertaining FGs and their counterparts.
We confine the number of parameters to formulate
the baryonic physics to the bare minimum, by
focusing only on the most important known processes and aiming for physically motivated implementations.
Our model is less complicated
than the semi-analytic models used previously for the study of FGs \citep{dariush2007, diazgimenez2011,gozaliasl2014,kanagusuku2016}
or the computationally costly treatments of star formation involving hydrodynamics \citep{cui2011, kundert2017}. \numbered{9}{Using this model, we reproduce (see Section \ref{sec:selection}) a
relatively weak connection between magnitude gap and group age found in i.e. \citealt{dariush2010} and \citealt{raouf2014}.}

We now describe the key elements of our model.


\paragraph{Normalization of star formation rate} We parametrize the halo-dependent star formation rate (SFR)
in terms of the halo mass and the halo's dynamical time, and set the overall normalization with a multiplicative constant $c_*$. That is,
\begin{equation}
\dot{M}_* = c_* H(M_h) K(M_h) M_h / t_\mathrm{dyn},
\end{equation}
\noindent 
where $M_h$ is halo mass,
$t_\mathrm{dyn}$ the dynamical time of the halo.
Here, 
two functions,
 $H(M_h)$ and $K(M_h)$, have been inserted to treat the physics of cold and hot gas accretion onto halos and supernova (and photoheating) feedback effects,
 respectively,
 as described next.

\paragraph{Hot-Cold Accretion Dichotomy} We represent the hot-cold accretion dichotomy  
with a characteristic transition halo mass $M_c$ such that star formation in halos with masses $M_h > M_c$ 
is quenched due to gravitational heating. We implement this with a smooth step function in terms of the log halo mass,
\begin{equation} \label{eq:step}
H(x) = \frac{1}{1+e^{-kx}}.
\end{equation}
We use $x = \log_{10} M_h / M_c$ and $\log_{10} M_c/M_{\odot} = 12.5 + 0.1 (z-2)$ in our fiducial model, with a width $k = 1$, consistent with
results from detailed hydrodynamic simulations (\cite{keres2009} ).
Thus, we use two parameters to describe the hot-cold accretion dynamics.

\paragraph{Environmental Effect} Gas accretion and dynamics in galaxies in the vicinity of massive halos with hot, extended atmospheres are subject to twin effects due to ram-pressure stripping and cold gas supply shortage.
These environment effects are well known, seen in detail in cosmological hydrodynamic simulations.
We parameterize this effect, based on insight gained in hydro simulations (e.g., \citealt{cen2014}),
by suppressing star formation in galaxies located within three virial radii of 
halos with hot accretion, with 
the suppression equal to $H(x)$
in Eq (\ref{eq:step}), where the massive neighbor halo in this case is the most massive halo with hot atmosphere
that encloses the halo in question within the three virial radii of the former.
\numbered{10}{There is ample observational evidence for such shock heating effects around clusters extending to a few virial radii (i.e. \citealt{balogh1999, wetzel2012, bahe2012, haines2015}).}

\begin{figure*}
\plotone{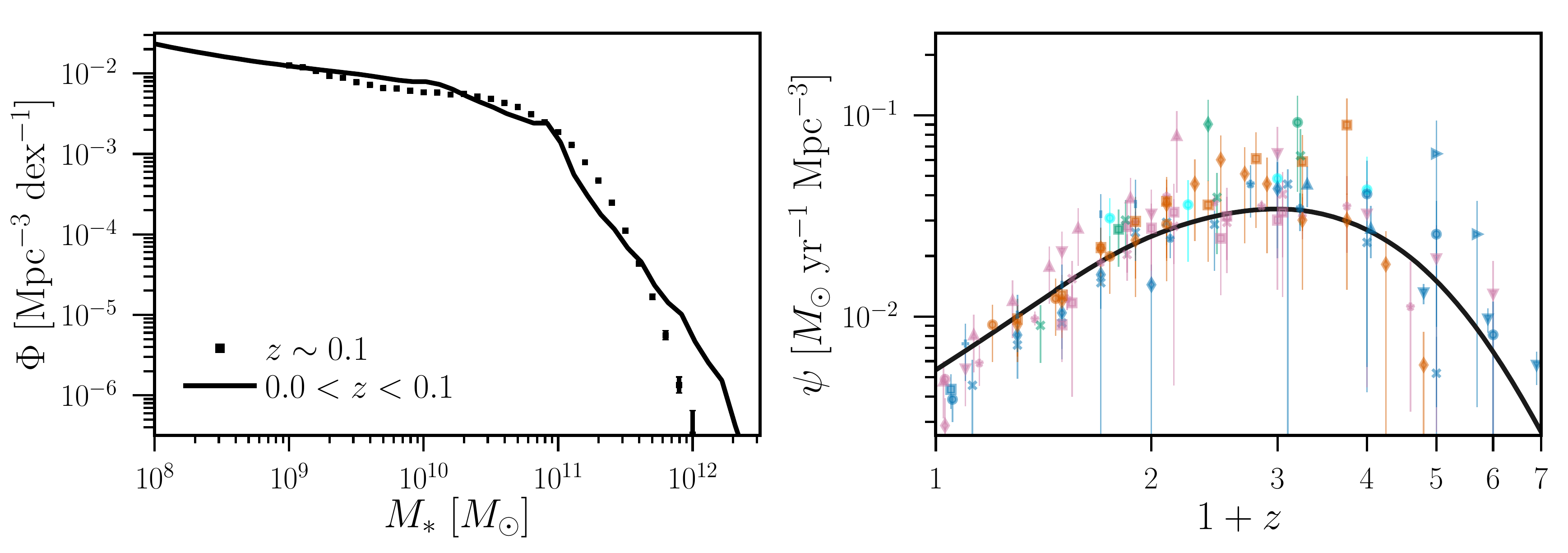}
\caption{
Left panel shows the stellar mass function at z=0.1 from observations
(black squares, \cite{moustakas2013}) and our model (black curve).
Right panel shows the SFR density as a function of redshift (Madau plot)
from observations (symbols) and our model (black curve).
Observational data sources and their symbols are as follows.
{\bf UV (blue)}:
\cite{2005Wyder} ($\bullet$),
\cite{2005Schiminovich} ($\blacklozenge$),
\cite{2011Robotham} ($\blacksquare$),
\cite{2012Cucciati} ($\times$),
\cite{2007Dahlen} ($\bigstar$),
\cite{2009Reddy} ($\blacktriangle$),
\cite{2012Bouwens,2012bBouwens} ($\blacktriangledown$),
\cite{2013Schenker} ($\blacktriangleleft$),
\cite{2006Yoshida} ($\blacktriangleright$),
\cite{2007Salim} ($+$),
\cite{2011bLy} ($\blacklozenge$),
\cite{2010vanderBurg} ($\bullet$),
\cite{2007Zheng} ($|$).
{\bf IR (pink)}:
\cite{2003Sanders} ($\bullet$),
\cite{2003Takeuchi} ($\blacklozenge$),
\cite{2011Magnelli} ($\blacksquare$), 
\cite{2013Magnelli} ($\times$),
\cite{2013Gruppioni} ($\bigstar$),
\cite{2010Rujopakarn} ($\blacktriangle$),
\cite{2009LeBorgne} ($\blacktriangledown$);
{\bf H-$\alpha$ (green)}:
\cite{2011Tadaki} ($\bullet$),
\cite{2009Shim} ($\blacklozenge$),
\cite{2011Ly} ($\blacksquare$),
\cite{2013Sobral} ($\times$).
{\bf UV+IR (cyan)}:
\cite{2010Kajisawa} ($\bullet$);
{\bf Radio (orange)}: 
\cite{2009Smolvic} ($\bullet$),
\cite{2009Dunne} ($\blacklozenge$),
\cite{2011Karim} ($\blacksquare$).
\label{fig:smf_and_madau}}
\end{figure*}

\paragraph{Supernova Feedback}  
We model supernova feedback 
using a redshift-dependent velocity dispersion cutoff $\sigma_c(z)$, below which star formation is suppressed. We use a smooth step function as in Equation \ref{eq:step} with $x = \sigma_\mathrm{v}/[(1+z)\mathrm{km/s}] - 25$, 
where $\sigma_v$ is 1-d velocity dispersion of the halo, and $k=0.2$.
Here, we also use two parameters to describe the collective feedback due to supernovae.
We considered adding feedback from photoheating of the intergalactic medium due to
reionization, which is relevant at high redshift.
We found that the negative feedback effect due to photoheating is subdominant to that
of supernova feedback based on simple physical considerations and hence is not included in our treatment.

In summary, we use a total of five parameters (one for overall normalization of star formation efficiency, two for cold-hot accretion and environmental effects,
and two for supernova feedback effects).
We implicitly assume that during mergers, the descendant halo receives all of the stars of the merging halos, as the contribution of the ejected stars, seen for example as intra-cluster light mounts
up to order of 10-20\%, still smaller fractions in groups and poorer environments
\citep{krick2007}.
\redout{Finally, we point out that we do not include AGN feedback,
in part because the physics is not well understood. 
A possible parameterization of AGN feedback where the negative effect is proportional to the stellar mass may not only
lead to degeneracies but also a circular situation with respect
to stellar masses. On the other hand, if one parameterizes 
AGN feedback to depend on the halo mass, it can be absorbed
into our two-mode accretion parameterization above.}
\numbered{3, 11}{We apply a simple suppression of star formation at high halo mass to match observations. This could physically arise from processes such as AGN feedback or gravitational shock heating, but we keep the prescription simple due to the relatively poorly understood underlying physics.}

\paragraph{Optimizing Physics Modeling Parameters}
The parameters in our star formation model, as detailed above, are then optimized
by matching two key observables -- the star formation history of the universe
and the galaxy stellar mass function at $z\sim 0$, shown 
in the right and left panels of Figure (\ref{fig:smf_and_madau}), respectively. 
Despite the small number of parameters employed, we obtain acceptable agreement with observations without sacrificing understanding.


\subsection{Definition of Fossil Groups} \label{sec:Definition}

\begin{figure}
\plotone{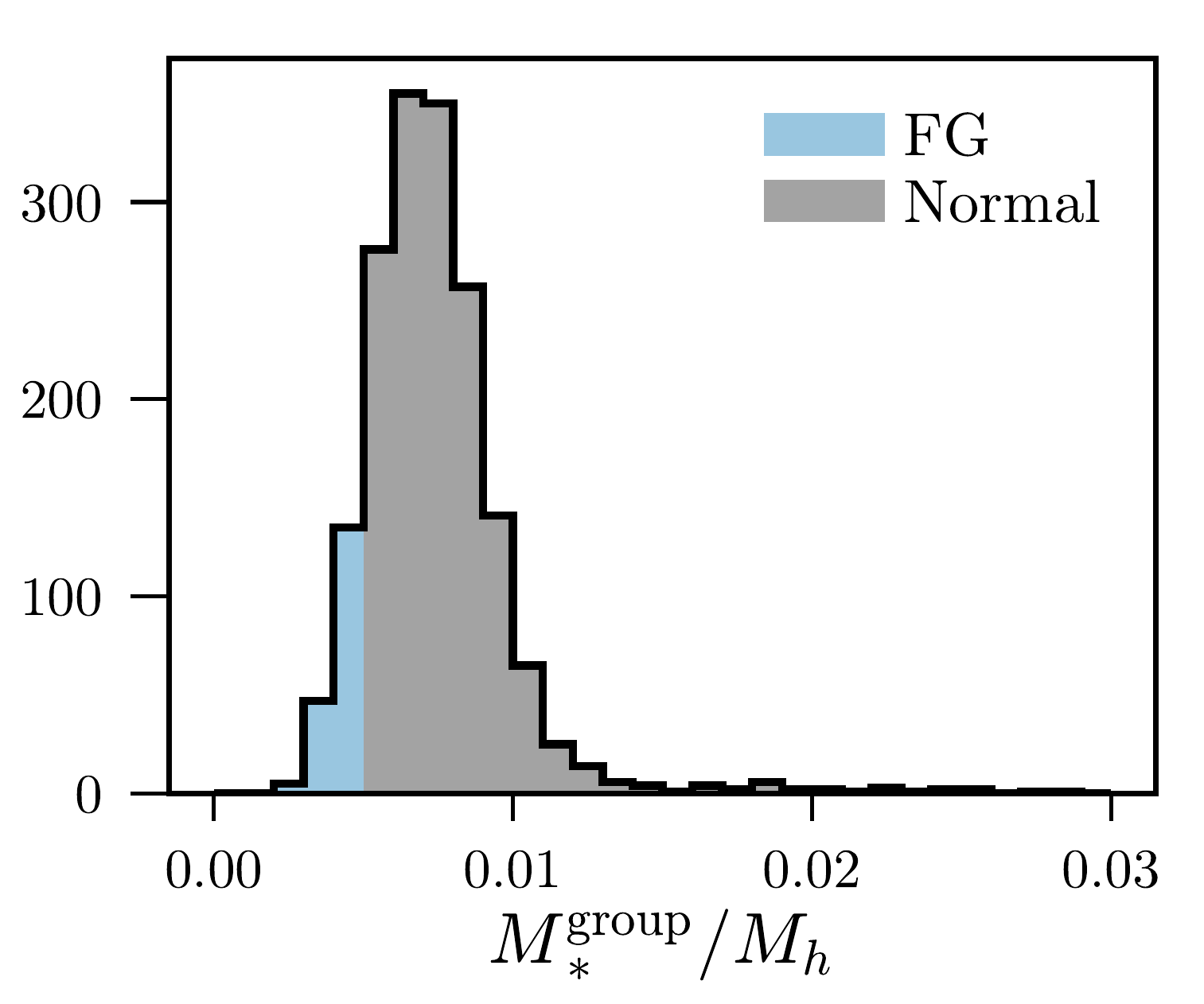}
\caption{
The probability distribution function of stellar to halo ratios
of groups of halo masses in the range $10^{13.4} M_{\odot} < M_h < 10^{13.6} M_{\odot}$.
Our standard method to select out low stellar content, FGs is shown by the cyan portion of the distribution, corresponding to a fraction is $\sim 10\%$ of all groups in the indicated halo mass range. 
\label{fig:FG_select}}
\end{figure}

\begin{figure}
\plotone{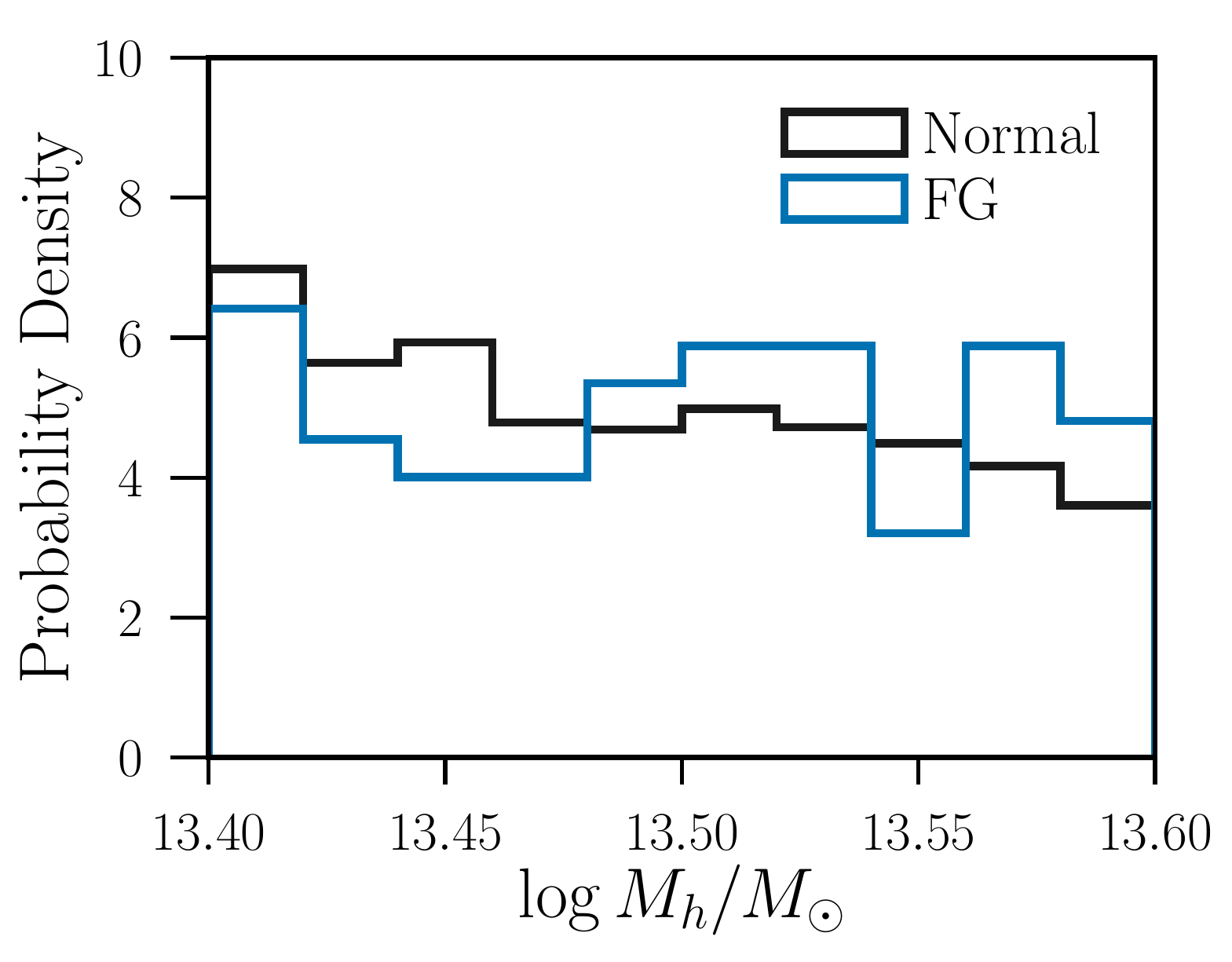}
\caption{The halo mass probability distributions resulting from our selection. 
Within this narrow mass range used for the analysis ($13.4 < \log_{10} M_h / M_{\odot} < 13.6$). \label{FG_halo}}
\end{figure}

\numbered{13}{We define fossil groups in this paper based on the ratio of stellar mass to halo mass. 
Compared to the various fossil identification methods based on magnitude gaps between the brightest galaxy and its
satellites, this definition is less sensitive to
the timing within the merger process that is ubiquitous in the hierarchical growth process of halos.}
Let $\langle M_*^{\group}/M_h\rangle$ be the ratio of total stellar mass to halo mass, averaged over all groups with halo mass $M_h=10^{13.4}-10^{13.6}M_\odot$ at $z=0$. 
A FG is defined to satisfy the following equation:
\begin{equation}
M_*^{\group}/M_h < 0.5 \langle M_*^{\group}/M_h\rangle.\label{eq:criteria}
\end{equation}
This definition, in addition to having the advantage of being relatively immune to the exact timing of merger processes
that may render the ranking based definition unstable,
is motivated by our desire of making the comparative statements robust,
not subject to uncertainties in the absolute values of the multiplicative parameters, such as $c_*$, in the model.
Although the halo mass range of real FGs may be larger than adopted here,
we choose a narrower range of halo masses to avoid possible mass segregation effects of the two classes
of groups, to more cleanly gain physical insight.
We define the ``normal" set to be the remaining groups in the same halo mass range $10^{13.4} M_{\odot} < M_h < 10^{13.6}$,
meeting the condition
\begin{equation}
M_*^{\group}/M_h \ge  0.5 \langle M_*^{\group}/M_h\rangle.\label{eq:normal_criteria}
\end{equation}
\noindent
We show these two samples in Figure \ref{fig:FG_select} with respect to their stellar mass
distribution, \numbered{14}{and highlight our selection of the most stellar-mass-deficient groups. For the selected mass range $M_h=10^{13.4}-10^{13.6}M_\odot$ at $z=0$, the overall distribution has a mean stellar-to-halo mass ratio of $\sim 0.007$ with a standard deviation of $\sim 0.002$.}
Figure \ref{FG_halo} shows the halo mass distributions of the two groups 
are similar.

To facilitate an assessment of selection methods based on luminosity gaps between ranking galaxy members
in a group, 
we use the ratio in stellar masses
as a proxy for the optical magnitude gap:
\begin{equation}
\Delta m_{12} = 2.5  \log_{10} \frac{ (M_*)_1 }{ (M_*)_2 }. \label{eq:def_m12}
\end{equation}
where $(M_*)_1$ is the stellar mass of the brightest member, and $(M_*)_2$ is the stellar mass of the second most massive galaxy in the group.
\numbered{17}{We restrict the second most massive member to be within $0.5 R_{\text{vir}}$.}
Since the most massive galaxies in groups are typically dominated by old stars,
the stellar mass is expected to be a reasonable proxy for luminosity used observationally.
\redout{
Perhaps, a more appropriate statement is that the observationally adopted luminosity gas measure
is a good proxy for the stellar mass gap for galaxies dominated by old stellar populations that is a fundamental quantity.}

\bigskip

\section{Results}

\subsection{Overall Star Formation Histories} \label{sec:starpoor}

\begin{figure}
\epsscale{1.1}
\plotone{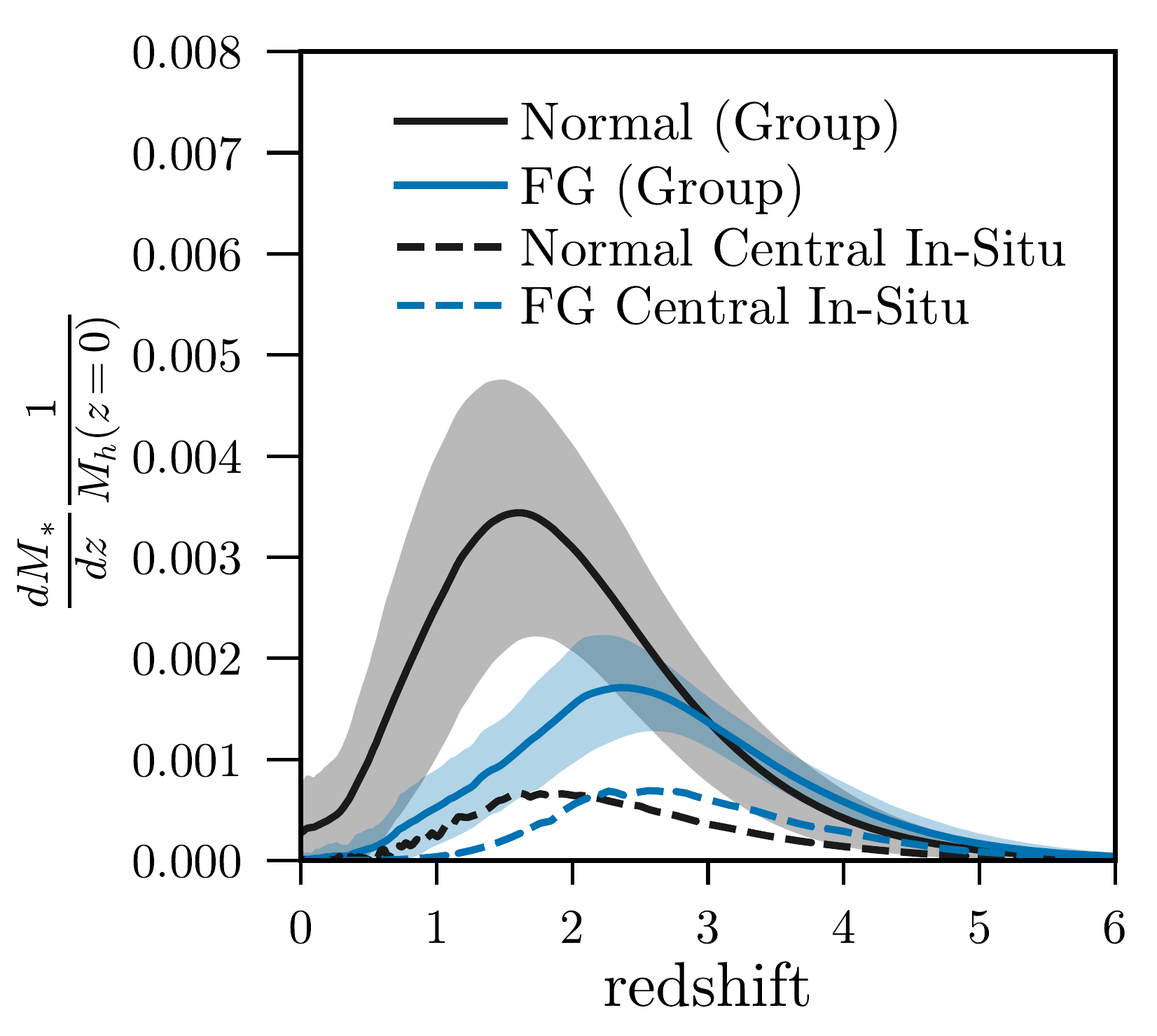}
\caption{
The formation rate of stellar mass per unit redshift 
as a function of redshift, $dM_*/dz$, divided by the final halo mass at $z=0$,
for the FGs (solid blue with shaded region) and normal (solid black with shaded region) 
samples with halo mass in the range of $10^{13.4} < M_h / M_{\odot} < 10^{13.6}$. 
We also show the same of the central galaxies for the two samples 
as dashed curves of matching colors.
For each sample the shaded region represents the $1\sigma$ range.
\label{stellar_pop_history}}
\end{figure}

FGs and normal groups show very different star formation histories in our model, with the primary differentiator being a suppression in FGs of the overall group star formation rate at low redshift.
Figure \ref{stellar_pop_history} shows the \numbered{18}{star formation histories of 
FGs and normal groups. We show the growth of stellar mass for the entire group (solid), and compare this to the stellar mass formed in-situ within the central galaxy (dashed).}
Both categories share a similar early history of star formation
down to about $z\sim 3$, at which point their SFR start to deviate from one another.
The SFR per unit redshift in 
FGs peak at a relatively high redshift, at $z\sim 2.5$,
whereas SFR per unit redshift in normal groups 
continues to rise until peaking at $z\sim 1.7$.
These trends in the SFR history of the central galaxies (dashed) follow that of the entire group for both cases.
With respect to the overall amplitude of SFR, 
a significant difference between FGs and normal groups is seen:
while in both cases the star formation histories of central galaxy and the entire group share a similar temporal shape and amplitude, the mean ratio of star formation in the central vs. the group at the peak of group star formation \numbered{19}{is about $0.4 \pm 0.1$ in fossil groups, but $0.2 \pm 0.1$ in normal groups.}

Consequently,
the FG sample, on average, tends to harbor a central halo which is more massive in stellar mass 
that the normal groups at all redshifts.
Since the star formation in FGs is considerably suppressed at the time of peak global star formation, the FGs tend to have lower stellar content.
We now examine the underlying reasons for these differences between 
the two groups.

\subsection{Shifted Self-Similar Halo Assembly Histories Between FGs and Normal Groups} \label{sec:starpoor}

\begin{figure}
\plotone{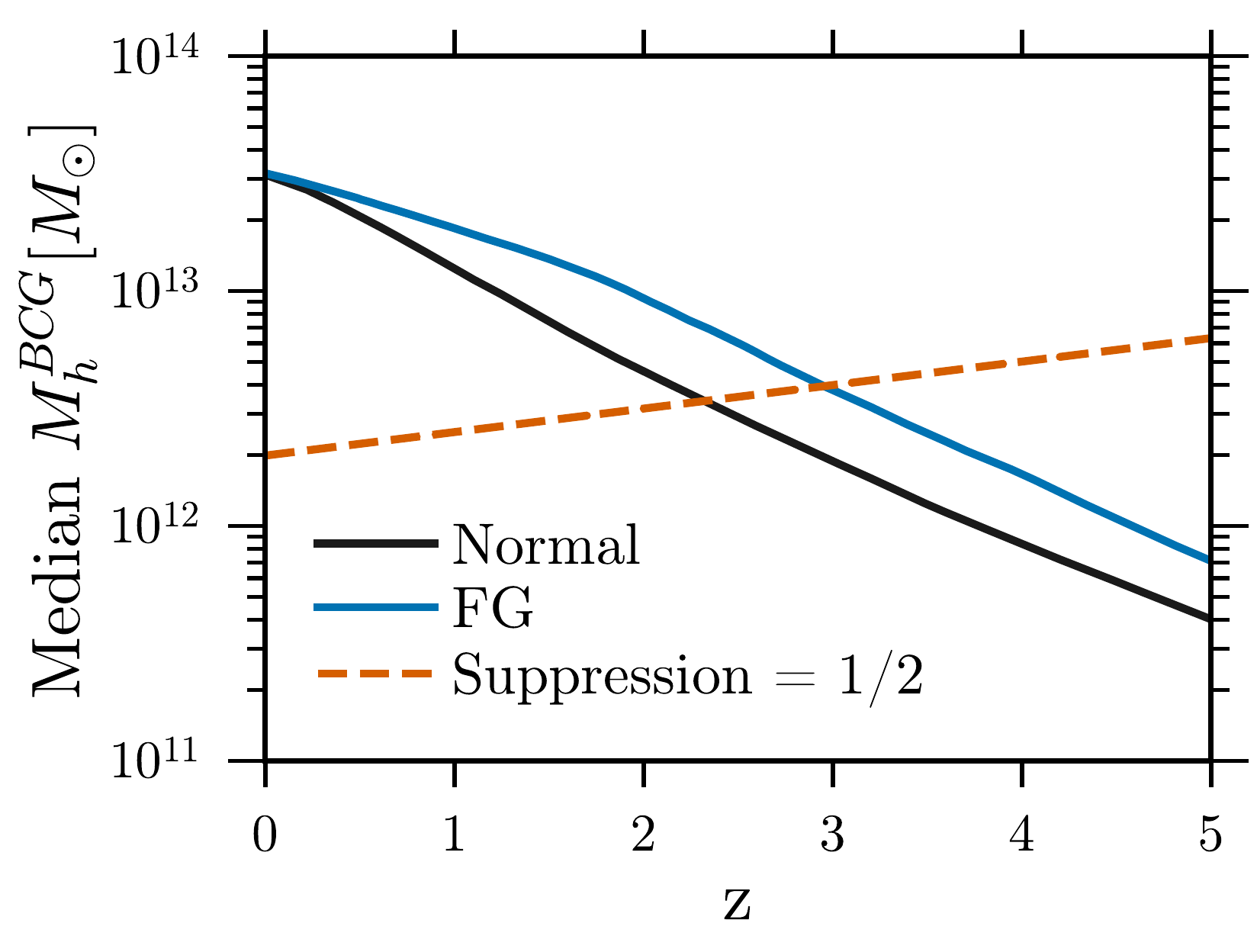}
\caption{The median mass of the central halos of FGs (solid blue) and normal groups (solid black)
as a function of redshift. 
Also shown as the dotted orange curve is the transition halo mass
between cold and hot accretion such that star formation is suppressed by an 1/2.
\label{BCG_growth}}
\end{figure}


\begin{figure*}
\plotone{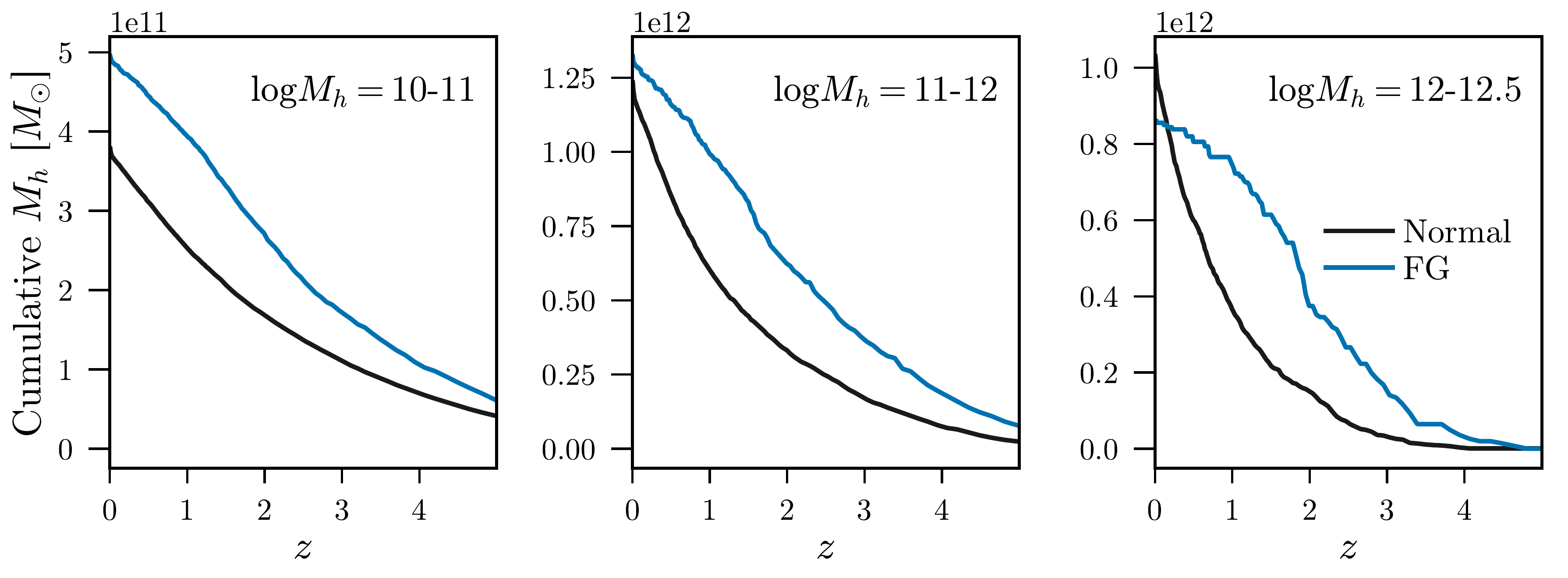}
\plotone{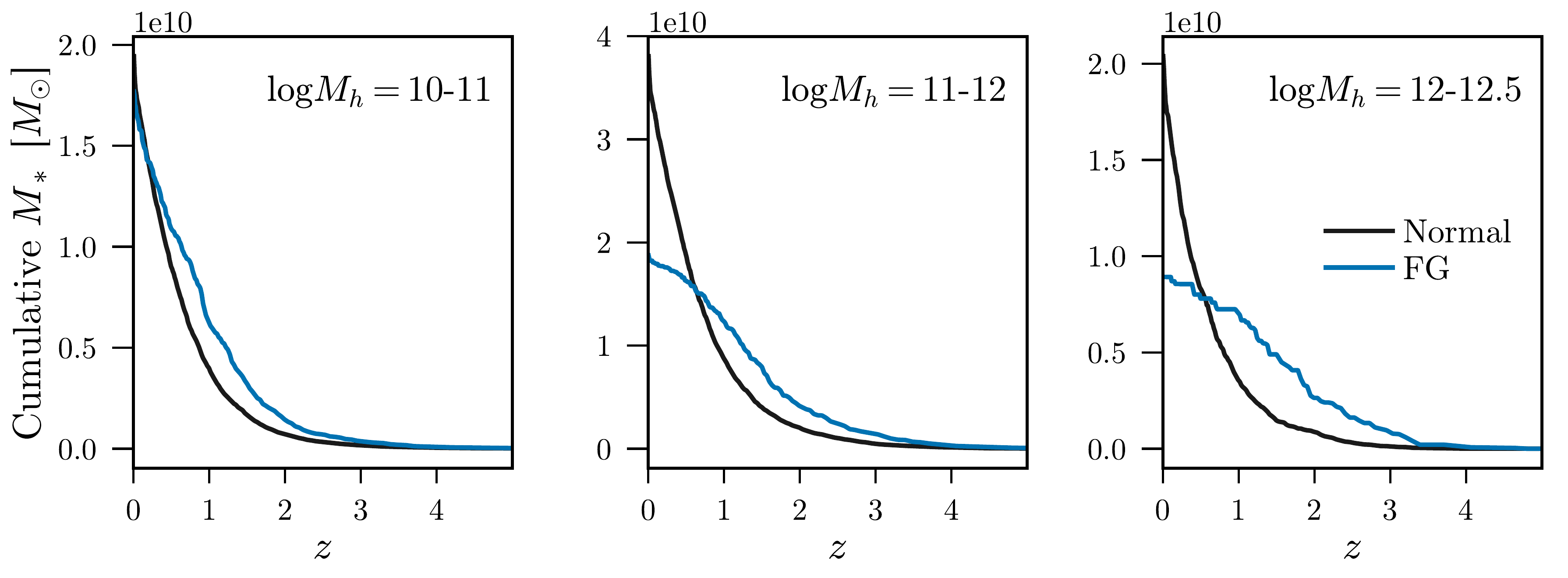}
\caption{The top row shows
the cumulative dark matter mass from mergers onto the progenitor, of halos 
in three different mass bins, 
$\log M_h/\msun = 10-11$ (left panel),
$\log M_h/\msun = 11-12$ (middle panel) and
$\log M_h/\msun = 12-12.5$ (right panel),
for FGs (blue dashed curves) and normal groups (black solid curves), respectively.
The bottom row shows the corresponding cumulative stellar mass.
\label{merger_over_redshift}
}
\end{figure*}

\begin{figure*}
\plotone{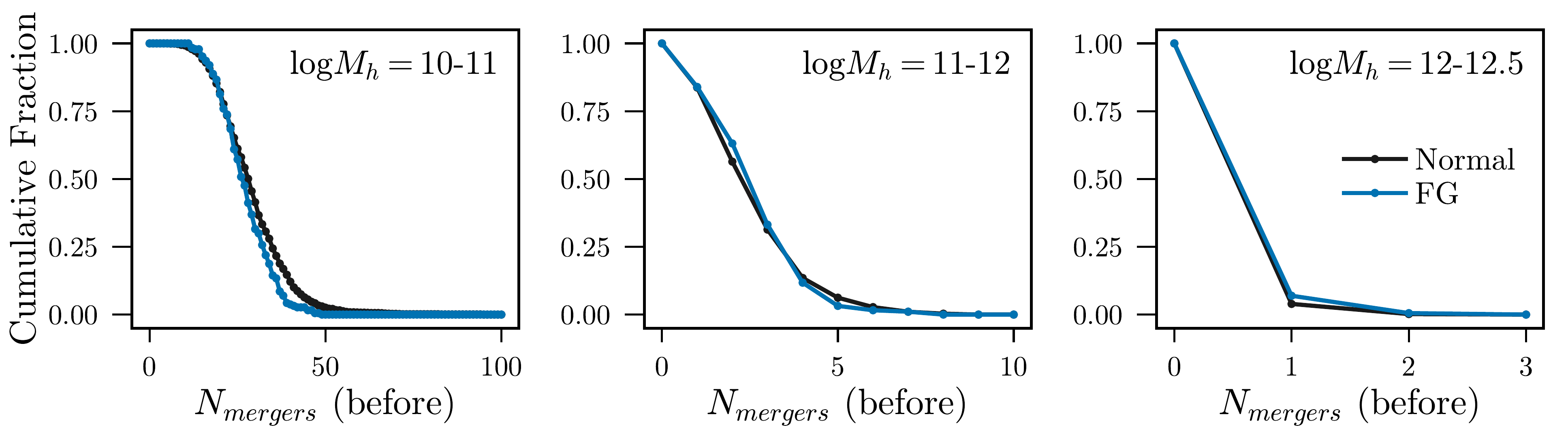}
\plotone{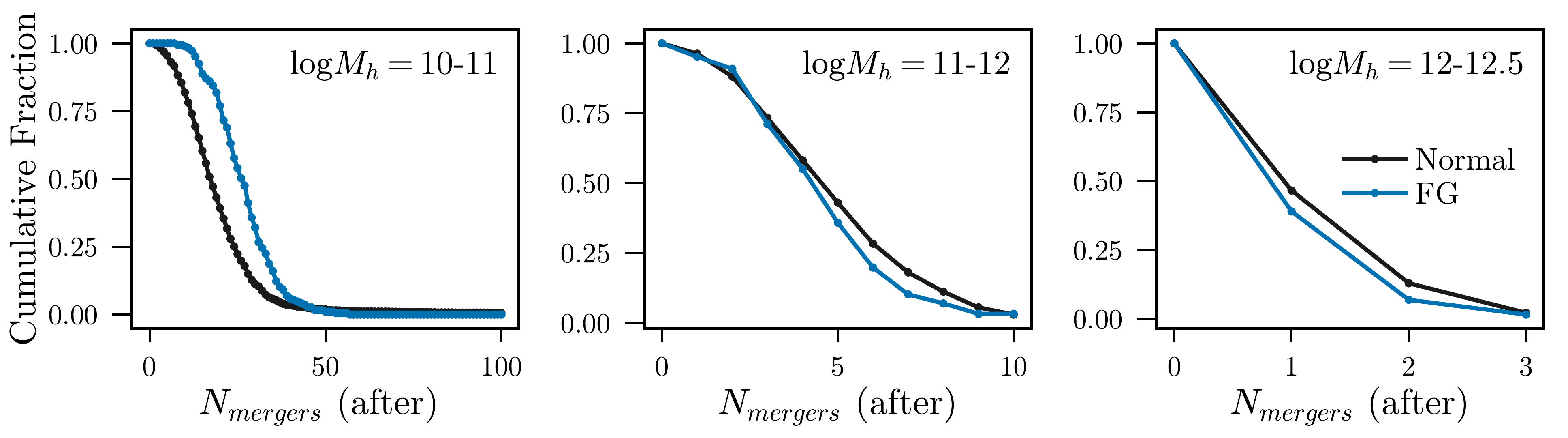}
\caption{Cumulative merger counts before first crossing $M_c$ for normal and FG.
\label{merger_fraction_over_redshift}}
\end{figure*}

The FGs have halo assembly histories which are shifted in time compared to the normal groups.
\numbered{21}{Recall that we impose a transition halo mass $M_c$ separating cold and hot accretion within our simplified star formation model.}
Figure \ref{BCG_growth} shows 
the median mass of the central halos of FGs (solid blue) and normal groups (solid black)
as a function of redshift, \numbered{22}{along with a measure of the suppression of star formation at high halo mass in our model. We indicate the halo mass at which star formation is suppressed by 
by a factor of two due to $M_c$.}
We see that the redshift at the crossing of the halo mass curves and $M_c$ curve (orange horizontal dashed line) 
in Figure \ref{BCG_growth} is conformal, direction-wise,
to the peak redshifts of star formation shown in Figure \ref{stellar_pop_history},
for both FGs and normal groups.
\numbered{22}{The FG and normal groups cross the halo mass where star formation is suppressed by a factor of 2 at $z\sim 3$ for FGs versus $z\sim 2.3$ for normal groups. This is part of the cause for a similar difference in the redshift of peak star formation, which is  $z\sim 2.5$ for FGs and $z\sim 1.7$ for normal groups. The lag between the redshift where substantial suppression starts and the peak redshift 
is because, while quenching occurs in the central halo, star formation may continue in satellites.}
We return to this point in \ref{subsec:conformity}.
Overall, 
the assembly histories of FGs clearly differ from normal groups in terms 
of the assembly redshift, despite our selection for the same halo masses at $z=0$.

The FGs also have satellite halos which follow their central halos in terms of earlier assembly histories.
In the top \numbered{24}{panels} of Figure \ref{merger_over_redshift} we show the redshift distribution 
of accreted halos in three different mass bins.
We see that there is a clear shift to higher redshift (i.e., earlier accretion) for the most
massive bins, $M_h=10^{12}-10^{12.5}\msun$,
for FGs than normal groups, but the overall mass accreted in this halo mass bin
is ultimately larger (by only $\sim 10$\%) for normal groups than for FGs.
For the intermediate halo mass bin,
$M_h=10^{11}-10^{12}\msun$,
a similar shift is seen but the two groups accumulate to about the same amount by $z=0$.
Finally, for the lowest mass bin,
$M_h=10^{10}-10^{11}\msun$,
the shift is consistent and goes all the way to $z=0$.
This examination indicates that 
the earlier assembly of the the FG is shared by the satellite halos that are accreted
by the respective main halos, over a large range of satellite masses.
We shall call this phenomenon ``halo assembly conformity".

The conspicuous difference between the redshift and amplitude of SFR peaks
between FGs and normal groups seen 
in Figure \ref{stellar_pop_history} raises the obvious question:
do FGs and normal groups acquire halos subhalos in a self-similar fashion
with respect to the peak redshift?
In other words, we ask whether the gravitational dynamics of FGs and 
normal groups are self-similar or not.
To address this question, we track the number of subhalos accreted 
before and after the respective redshift peak of $M_c$ crossing.
In the top row of Figure \ref{merger_fraction_over_redshift} 
we show the histograms of the number
of accreted satellite galaxies in three mass bins for FGs (blue curves) and normal groups (black curves)
from high redshift down to the redshift for an individual group reaching $M_c$.
The bottom row of Figure \ref{merger_fraction_over_redshift} is similar to the top row but 
for the redshift range of $z=0$ to $z=z_{peak}$.

In both redshift ranges, demarcated by the respective redshift peak,
no statistically significant differences are seen between FGs and normal groups
with respect to the mass distribution of satellite halos.
From this we can conclude that
the gravitational dynamics alone, i.e., the number of halos or the type of halos accreted,
is not a direct cause 
for the difference in the amplitude and redshift of the star formation
peak between FGs and normal groups,
if the stellar content of a galaxy depends only on the halo mass.
Thus, the difference in stellar masses between FGs and normal groups 
ought to be rooted in the difference in stellar mass to halo mass ratios 
between those comprising FGs and those comprising normal groups.

\subsection{Origin of Stellar Content Difference between FGs and Normal Groups} 
\label{sec:stellarcontent}

\subsubsection{Baryonic Environment Effects} 

We argue that quenching effects coming from the environments of satellites do not play a major role in the stellar content differences between FGs and normal groups.
FGs and normal groups both experience some degree of environmental quenching \numbered{26}{in our model, when the central halo mass grows more massive than $M_c$ and enters
the hot accretion regime. We apply to the bound subhalos the same suppression factor that the more massive central halo experiences, to model an environmental suppression by the central halo on star formation in the satellites.}
We impose this same suppression factor on every satellite 
halo within three virial radii based on hydrodynamic simulations \citep{cen2014}. 
Since the central halos of FGs become more massive at earlier times,
the difference in stellar content for FGs and normal groups at $z=0$ 
is then a result of both satellites and central progenitors of FGs 
have lower stellar content than their counterparts of normal groups. 
We argue in this section that these effects are subdominant, by comparing the stellar content of the same groups, using models with and without this effect.

\begin{table}[t]\begin{center}
\begin{tabular}{ccc}
\hline
 Type &  $z_{1/2}$ &  $z_{1/2}$ \\
  & merged & $z=0$ sat \\
\hline
FG (env)      & 1.1   & 1.7 \\
FG (no env)      & 0.9 & 1.4  \\
Normal (env)      & 0.45   & 0.63  \\
Normal  (no env)      & 0.43   & 0.65  \\
\hline
\end{tabular}
\end{center}

\caption{The effects of environmental suppression on mass evolution with redshift. We show stellar mass in progenitors which have already merged into the central ("merged") and stellar mass in progenitors of present-day group members which have not yet merged ("$z=0$ sat"). }
\label{tab:envz}
\end{table}

\begin{figure}
\plotone{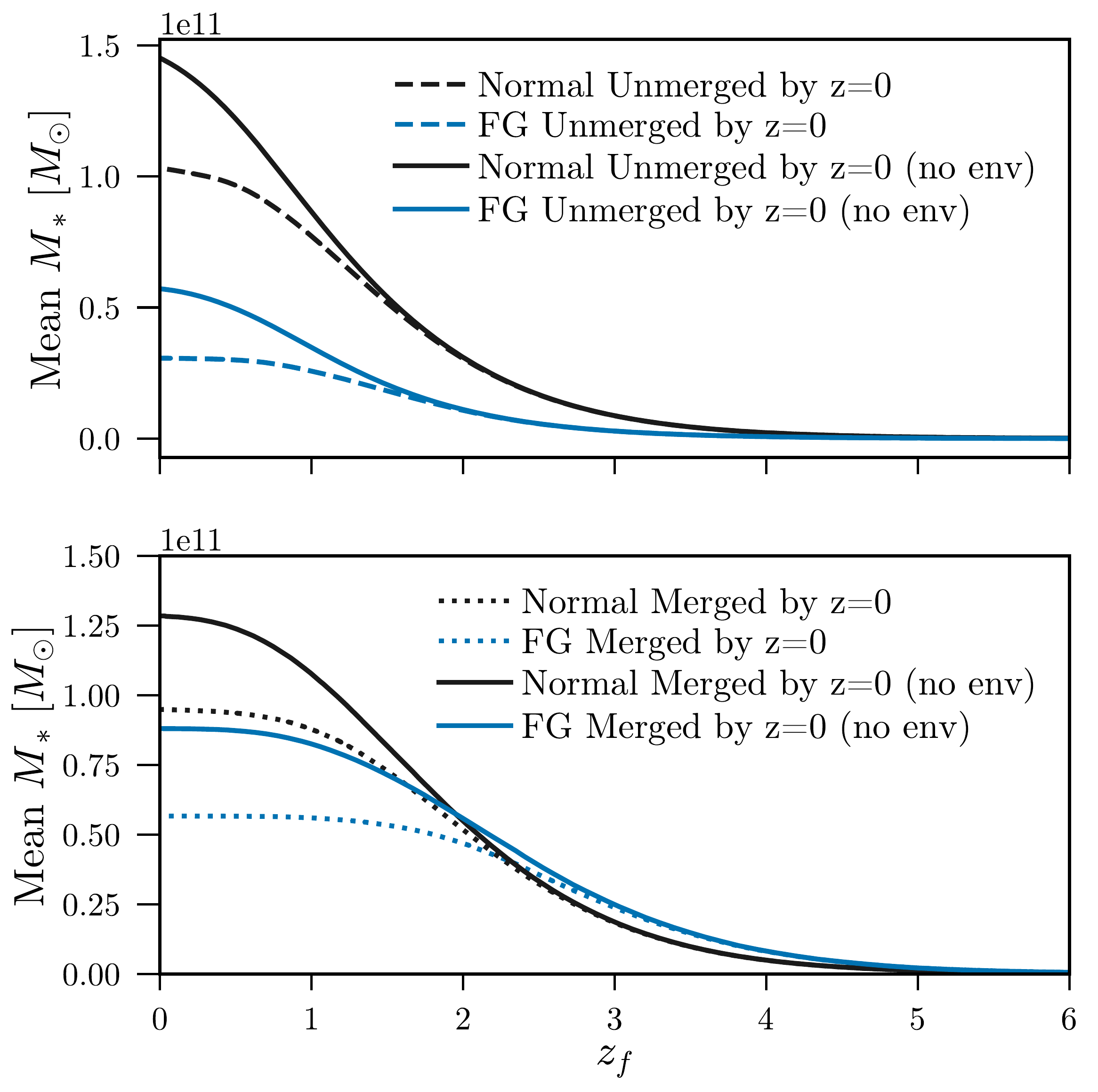}
\caption{We show the effect of the environmental suppression for both normal and FG. Summary statistics like the half-mass redshift are provided in Table \ref{tab:envz}. In the top panel, we show stellar mass in progenitors of present-day group members which have not yet merged. We show these for normal groups (black) and FG (blue), with environmental suppression (dashed) and without (solid). In the lower panel, we show the same quantities for progenitors of the central halo which have already merged by redshift 0. \label{fig:noenv}}
\end{figure}

We show in Figure \ref{fig:noenv} that the environment effect is significant but 
is unable to account for most of the difference
in the stellar content between FGs and normal groups.
\numbered{27}{The top panel of the figure shows the growth of stellar mass in progenitors of present-day group members which have not yet merged with the central halo. In both normal groups and FGs, the environmental suppression has a $\sim 20$\% impact on the final stellar mass. A similar magnitude is shown in the bottom panel, which shows the mass growth of stellar mass in progenitors which have already merged at $z \sim 0$. The normal groups and FGs both experience a similar amount of lost final stellar mass from mergers. }
This indicates
that environment effect is significant but does not constitute a major contributor to 
the difference of stellar content at $z=0$ between FGs and normal groups.
But this findings is in apparent contradiction to the redshift evolution
of stellar mass shown in Figures \ref{stellar_pop_history}, \ref{merger_over_redshift} and in Figure \ref{stellarmassform} below.
We continue looking for the main culprit and a self-consistent 
physical explanation.

\begin{figure*}
\plotone{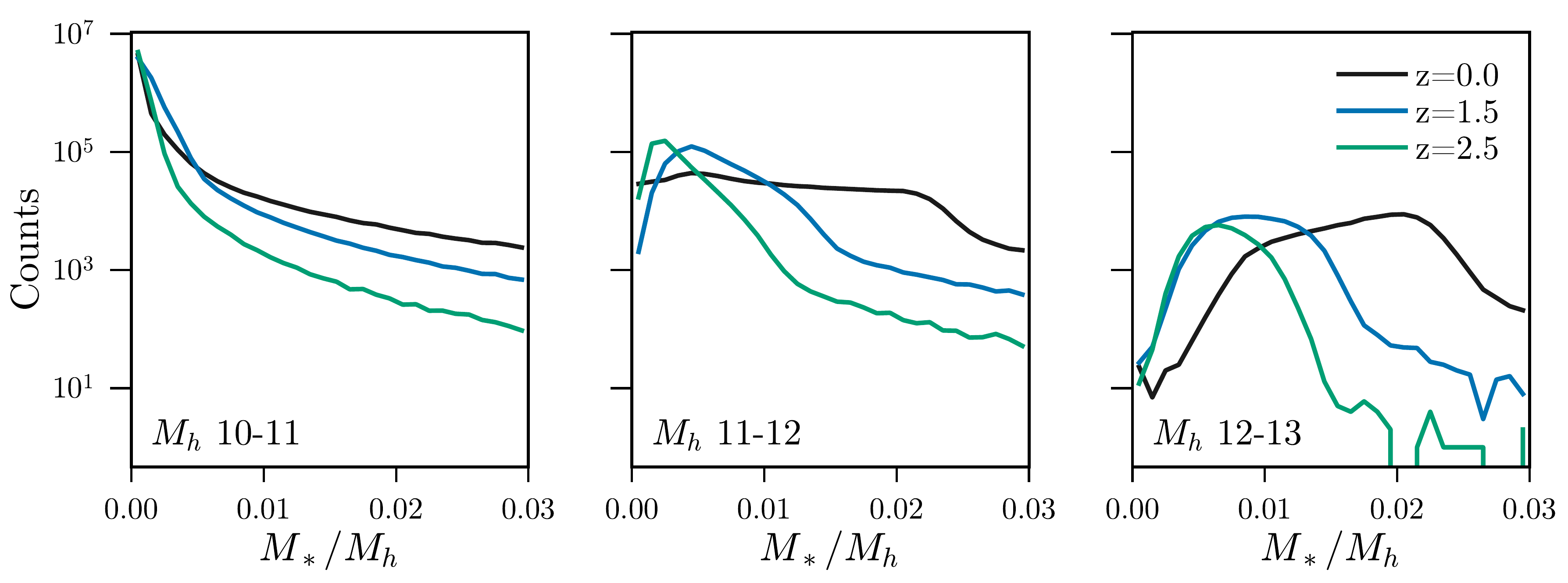}
\caption{The redshift evolution of the stellar-mass-to-halo-mass ratio for halos of 
$\log_{10}M_h=10-11$ (left), $11-12$ (middle) and $12-12.5$ (right).
In general, high redshift halos tend to be deficient in stellar mass. \numbered{29}{The global stellar-mass-to-halo-mass ratio evolves from 0.005 to 0.01 and then 0.02 at $z=2.5$, 1.5 and 0.0, respectively, shown.} \label{richness_evolution}}
\end{figure*}

\subsubsection{Effects due to Central-Satellite Halo Assembly Conformity}  \label{subsec:conformity}

The earlier assembly of main halos for FGs is paralleled, in time,
by a corresponding stellar component
in an increasingly amplified way with increasing subhalo masses.
In the bottom row of Figure \ref{merger_over_redshift} we show the redshift distribution 
of accreted stellar mass from the halos for three different mass bins.
A differential behavior for halos with different masses 
is visible:
the ratio of the cumulative stellar mass of FGs to that of normal groups is seen to 
increase with increasing halo mass of the subhalo.
This suggests decreased star formation rates in subhalos of higher \numbered{28}{masses}
of FGs relative to their counterparts of normal groups.
In other words, while stars in FGs form earlier and at a lower efficiency
than their counterparts in normal groups,
the difference increases with the halo mass of subhalos.

Figure \ref{richness_evolution} shows the redshift evolution of the stellar-mass-to-halo-mass ratio for halos in three $\log_{10}M_h$ ranges in three separate panels. 
A similar difference is seen for halos of mass in the range $10^{11}-10^{12}\msun$ (middle panel) 
and $10^{10}-10^{11}\msun$ (left panel), except that for the smallest halo mass bin 
the distributions at all redshifts peak close to zero, suggesting a
large number of starless subhalos.
In all cases, the trend of increasing stellar to halo mass ratio with decreasing redshift
continue to $z=0$ (solid curves in all panels).
Thus, Figure \ref{richness_evolution} may be able to provide 
the reason why FGs are more stellar deficient than
the normal groups,
simply because progenitors of FGs mature (i.e., crossing $M_c$)
at a higher redshift than the normal groups, when the stellar to halo mass ratio
of all constituents accreted is generally lower by a factor of roughly two.
Observationally, it is found that high redshift galaxies are indeed more gas rich and relatively stellar poor at a given halo mass of relevance \citep[e.g.,][]{2013Carilli}, consistent with our model results.

\begin{figure}
\plotone{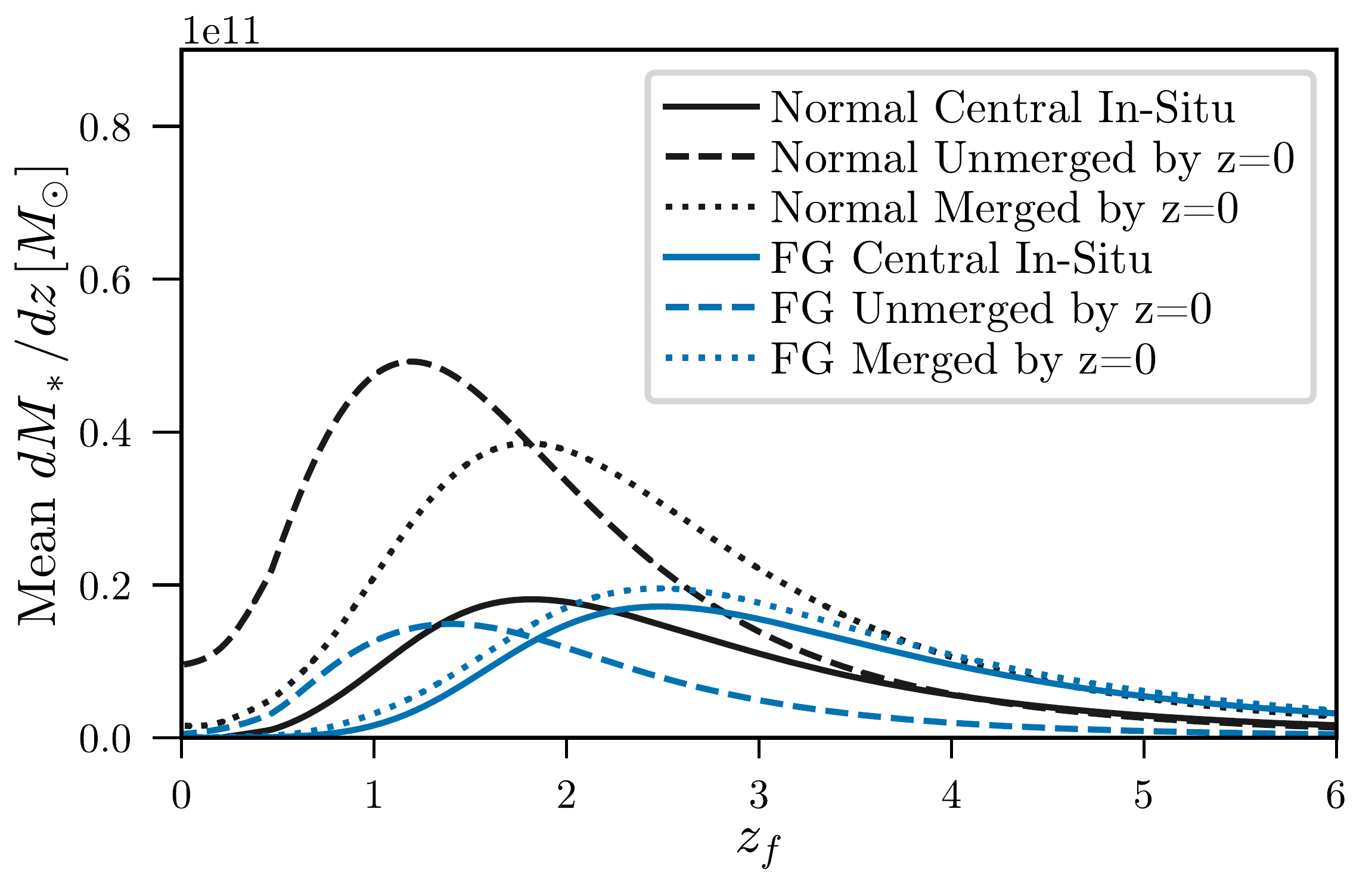}
\caption{\label{stellarmassform}
The stellar mass as a function of the \textbf{formation} redshift 
for three separate components: in-situ star formation in the main progenitor (solid curve),
accreted satellites that have merged into the central BCG (dotted curve)
and accreted satellites that have not merged into the central BCG (i.e.,
satellites of the \numbered{31}{BCG} at $z=0$, dashed curve),
for FGs (blue curves) and normal groups (black curves), respectively.
}
\end{figure}

\begin{figure}
\plotone{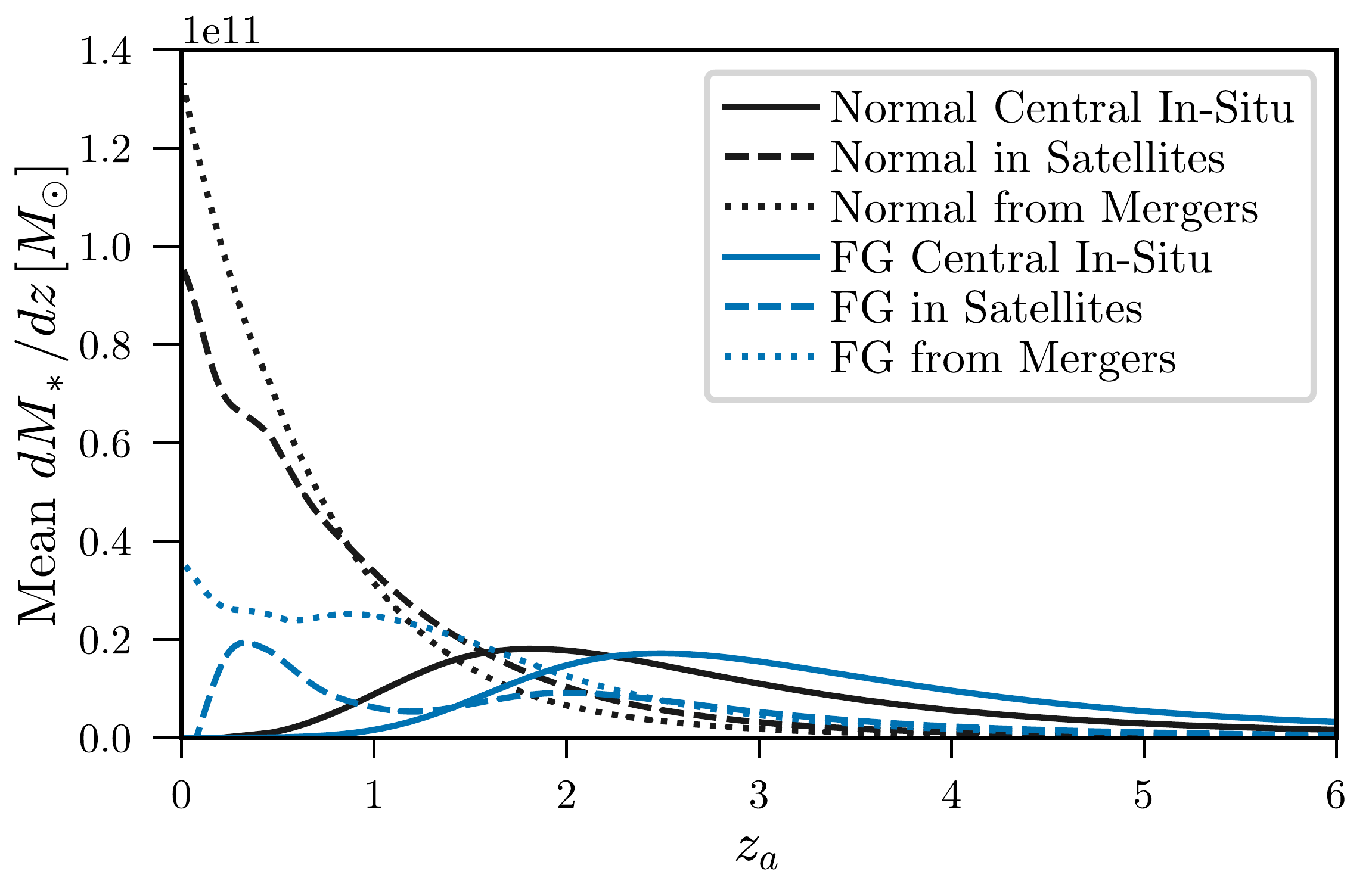}
\caption{\label{stellarmassaccretion}
The stellar mass as a function of the \textbf{accretion} redshift 
for three separate components: in-situ star formation in the main progenitor (solid curve),
accreted satellites that have merged into the central BCG (dotted curve)
and accreted satellites that have not merged into the central BCG (i.e.,
satellites of the BGC at $z=0$, dashed curve),
for FGs (blue curves) and normal groups (black curves), respectively.
}
\end{figure}

We argue that another part of the lower stellar content in fossil groups comes from this global relationship between redshift and richness demonstrated in Figure \ref{richness_evolution}, combined with the earlier satellite accretion histories of FGs compared to normal groups.
The stellar-mass-to-halo-mass ratio shown in 
Figure \ref{richness_evolution} is the ratio at the respective redshift, which may be different from the ratio at the time of accretion/acquisition
by the central galaxies.
To check this,
Figure \ref{stellarmassform}
shows the stellar mass accumulated as a function of the star formation
redshift classified into three types: in-situ star formation in the main progenitor (solid curve),
accreted from satellites that have merged into the central BCG by $z=0$ (dotted curve)
and accreted from satellites that have not merged into the central BCG by $z=0$ (i.e.,
satellites of the BGC at $z=0$, dashed curve),
separately for FGs (blue curves) and normal groups (black curves).

The star formation histories of both in-situ stars and ex-situ
stars of normal groups may be approximately obtained by 
shifting the peaks of their counterparts of FGs to lower redshifts by roughly 
$\Delta z\sim (1.2, 0.7, 0.1)$
for (in-situ, ex-situ merged, ex-situ unmerged), respectively.
In terms of star formation age, the three components in the order of increasing age
are (in-situ, ex-situ merged, ex-situ unmerged) for the FGs, with the ``in-situ" component 
being older than the ``ex-situ merged" only by about $\Delta z\sim 0.5$.
The ``ex-situ merged" component is older than the ``ex-situ unmerged" component
by $\Delta z\sim 1$ for FGs.
The situation for normal groups display quite different behaviors.
We see that the in-situ stars in normal groups are on average slightly younger than
the accreted merged stars but older than the accreted unmerged stars.
Consequently, in the BCGs of FGs there is an old-to-young
stellar age gradient from the center to the outskirt,
where the opposite is expected for normal groups.
Moreover, the age contrast between central BCGs and satellite galaxies in BCGs
is expected to be much larger than that for normal groups.
These predictions are important and should be testable observationally.

We find that the fossil groups have very different satellite histories.
Figure \ref{stellarmassaccretion}
shows the stellar mass as a function of the \textbf{accretion} redshift 
for three separate components: in-situ star formation in the main progenitor (solid curve),
accreted satellites that have merged into the central BCG (dotted curve)
and accreted satellites that have not merged into the central BCG (i.e.,
satellites of the BGC at $z=0$, dashed curve),
for FGs (blue curves) and normal groups (black curves), respectively.
The histories of the in-situ star formation rate,
as a function of formation redshift in Figure \ref{stellarmassform}, 
identical to that 
as a function of accretion redshift in Figure \ref{stellarmassaccretion},
for FGs and normal groups,
display similar shapes characterized by a gradual rise at early times \numbered{32}{($z > 2.5$) followed 
by a somewhat steeper decline past the peak towards lower redshift ($z < 2.5$).
The location of the in-situ star formation peak coincides with the time when
the central halo of the group enters hot accretion mode, with FGs peaking at $z \sim 2.5$ compared to a peak at $z \sim 2$ for normal groups.}

However, a comparison of dashed curves (unmerged satellites) or dotted curves (merged satellites)
between Figure \ref{stellarmassaccretion} 
and Figure \ref{stellarmassform} 
indicates that the formation redshift and accretion redshift of stars in satellites 
are very different.
In particular, while the formation redshift of satellites display a shape - early rise followed by late decline -
similar to that of the in-situ star formation, albeit with different peak locations,
the accretion redshift of stars in satellites have different, varied temporal profiles.
Third, perhaps profoundly,
the difference in accretion redshift distribution between FG satellites and normal group satellites
is large.
We see that, whereas the accretion rates of both merged and unmerged satellies in normal groups' satellites 
continue to rise until $z=0$, the accretion rate of merged satellites of FGs initially rise from $z\sim 2$
to $z\sim 1$ and then largely flat, only to be followed by a moderate rise from $z\sim 0.2$ to $z=0$,
and the accretion rate of unmerged satellites of FGs shows a very gradual rise from high redshift to 
$z\sim 2$ then declines to $z\sim 1.3$, followed by a rise to $z\sim 0.3$, after which it declines
until $z=0$. We have yet to fully understand the physical reasons behind 
the complexity of this behavior.




\begin{table}[t]\begin{center}
\begin{tabular}{ccc}
\hline
Model  &  $M_*^{\text{FG}}(z=0)$ & $M_*^{\text{normal}}(z=0)$  \\
  &  $[10^{11} M_{\odot}]$ & $[10^{11} M_{\odot}]$  \\
\hline
env      & 1.4 & 2.4  \\
no env      & 1.9 & 3.1  \\
\hline
\end{tabular}
\end{center}

\caption{The effects of environmental suppression on total group stellar mass at $z=0$. The star formation suppression in FGs arises $\sim$70\% from gravitational effects (the difference in the bottom row of this table) and $\sim$30\% from environmental effects (the difference in the left column).}
\label{tab:envMs}
\end{table}

\begin{table}[t]\begin{center}
\begin{tabular}{ccc}
\hline
 Model &  $M_*^{\text{FG}} / M_*^{\text{normal}}$ & $M_*^{\text{FG}} / M_*^{\text{normal}}$  \\
  & merged & $z=0$ sat \\
\hline
env      & 0.61  & 0.37  \\
no env      & 0.70 & 0.49  \\
\hline
\end{tabular}
\end{center}

\caption{The effects of environmental suppression on relative stellar mass deficiency in our fossil groups, from different mass sources. The "merged" column refers to stellar mass in the central, obtained only from mergers. The $z=0$ sat column refers to mass in satellites in the group at $z=0$.}
\label{tab:envMs_ratios}
\end{table}

\begin{figure*}[t]
\epsscale{0.8}
\plotone{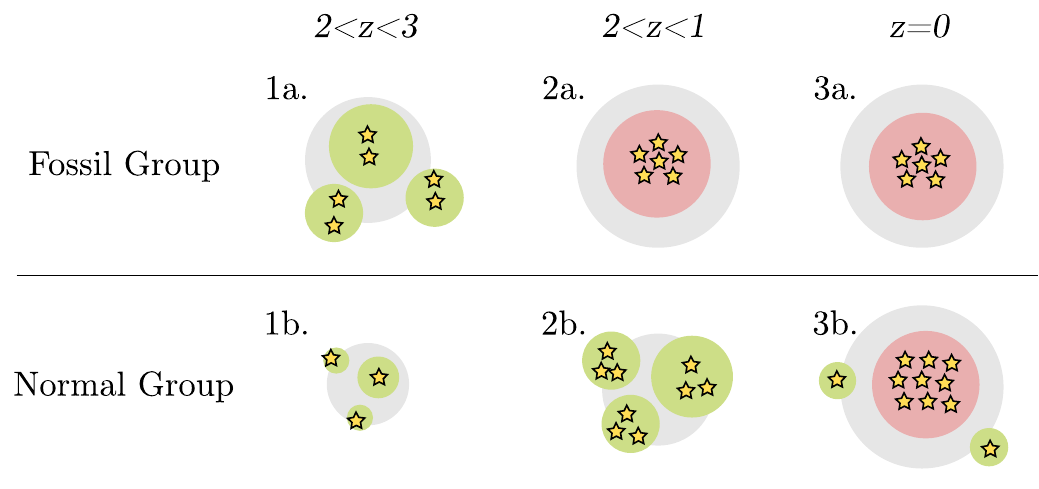}
\caption{\numbered{37 (updated figure)}{} A visual illustration of the essence of the key trend our model model predicts. 
The top and bottom rows are the star formation and halo growth history for FGs and typical groups 
of the same mass at $z=0$, respectively.
In a typical group (the bottom row), star-forming galaxies (green) begin to cluster at redshift $\sim 3$ (1a). They continue to accrete matter and form stars (1b) at redshifts $2 < z < 1$, until hierarchical merging leads to a local gravitational potential which heats the gas enough to quench star formation (1c). Our FGs also begin as star-forming galaxies at high redshift (2a), but assemble their halos earlier. Our FGs come from more concentrated initial density fluctuations, leading to a more massive central progenitor at redshift $\sim 2$ (2b). This quenches star formation early in the central halo, but also suppresses star formation in satellites within a few virial radii (shown as a dashed line). Star-forming galaxies in our model FGs tend to be consumed by the central halo earlier than in typical groups (2a to 2b), and such high-redshift galaxies tend to have lower stellar mass at fixed halo mass. The surviving satellites (2b) are quenched by heating of the environment by the central halo. The end result (2c) is a group-sized halo with the same halo mass, but has lower stellar mass.\label{cartoon}}
\end{figure*}


To summarize the complex trends in simple terms:
(1) The central galaxies and satellite galaxies in FGs formed relatively earlier than
their counterparts in normal groups.
(2) There is a general trend of decreasing stellar mass to halo mass ratio 
with increasing redshift at a given halo mass.
(3) The satellite galaxies in FGs get acquired by their central galaxies 
at  higher redshifts than their counterparts in normal groups.
(4) There is varying redshift intervals between the stellar assembly redshift
and accretion redshift of satellite galaxies in both FGs and normal groups.
Altogether, the differential between formation and accretion redshifts
and the differential of those between FGs and normal groups
ultimately contribute to the about 70\% of the difference
in stellar content between FGs and normal groups at $z=0$,
whereas baryonic environmental effects account for about 30\% of the difference.

Since the amount of dark matter mass in FG halos and normal groups 
is the same and since the number and mass distribution of halos accreted
before and after the central halo assembly redshift 
are the same for FGs and normal groups (see Figure \ref{merger_fraction_over_redshift}),
whatever physical effects responsible for the difference in stellar content
between FGs and normal groups must be then pertaining to baryonic effects
that affect star formation.
We ultimately ascribe the effect to declining star formation efficiency with decreasing
redshift from roughly at the peak to $z=0$
reflected in the temporal shape of the \numbered{33}{Madau plot (i.e. the right panel of Figure \ref{fig:smf_and_madau})}.
\redout{This explanation may at first appear to be paradoxical,
since FGs formed earlier may be expected to have a high star formation efficiency.
This is in fact a mis-perception. 
The star formation peak of the Madau plot itself
is not equal to a higher stellar content at a given halo mass.} \numbered{34}{Note that the Madau plot shows instantaneous star formation with respect to redshift, rather than time. Thus the peak of the Madau plot does not exactly correspond to either when the majority of star formation occurs, nor the time of peak star formation efficiency.
Thus, the Madau plot is in fact consistent with both 
an increasing star formation rate and a decreasing stellar mass to halo mass halo with increasing redshift at a given halo mass.}

\redout{We note another testable prediction of our model:} \numbered{35}{Our model predicts that the stellar mass within the ``unmerged'' satellite
component should display} a rapid drop from $z\sim 0.3$ to $z=0$ in FGs,
where this component in normal groups continues to rise rapidly towards $z=0$.
\numbered{35}{Thus, our model} says that the radial distribution of satellite galaxies in normal groups
is significantly more extended than that of the FGs.
\redout{This, we show,} \numbered{36}{This also} manifests in the difference in galaxy correlation functions
shown later.
Another possible testable prediction is also noted.
While the ``merged" satellite component is older than the ``unmerged" satellite component
in FGs, the former is accreted later than the latter to their respective locations.
This last point may be indicative that while there is continued
merging of satellites into the central BCGs of FGs,
the supply rate of satellites into the FGs (through the virial radii) is much lower at low redshift.

We now summarize the physical picture that has emerged from our analysis in a cartoon,
shown in Figure \ref{cartoon}.
We see that the FG halos assemble earlier than normal groups, as indicated by the larger 
satellite halos with more stars in the top left panel compared to the bottom left panel at $z\sim 2-3$.
By $z=1-2$ the FGs have largely matured, whereas the normal groups now just entered
their prime star formation and halo/stellar accretion peak, as indicated by the middle panels.
The most important point to notice is that the satellite halos of a same mass that is 
accreted by normal groups at $z=1-2$ are more stellar rich than those accreted by FGs at $z=2-3$.
By $z=0$, the cumulative difference in stelllar mass accreted from satellites and in-situ star formation
result in the difference in stellar mass between FGs and normal groups.

\subsection{Comparison with Magnitude Gap Methods} \label{sec:selection}

\begin{figure}
\plotone{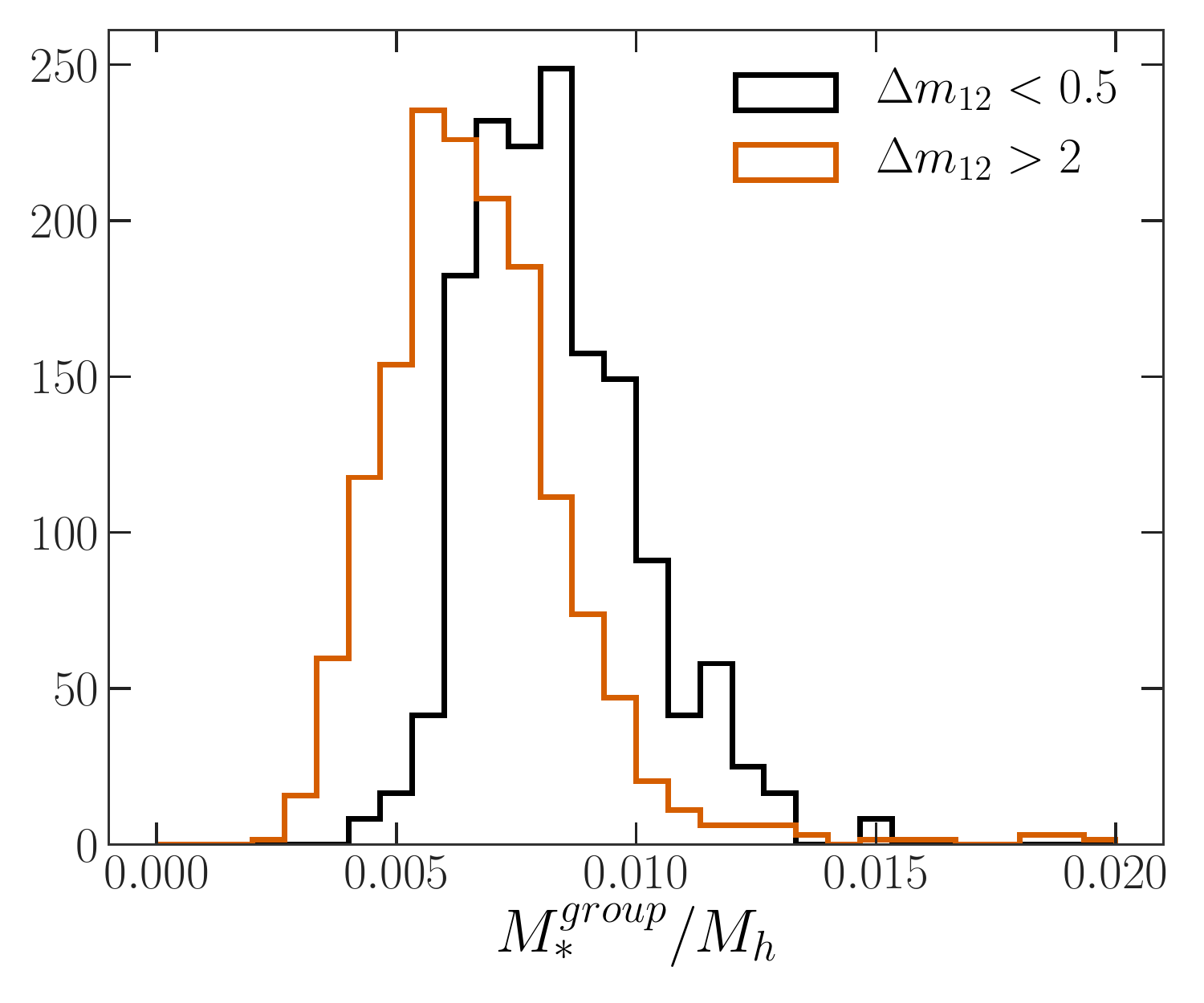}
\caption{
\numbered{17, 45 (updated figure)}{}
Two separate probability distribution function of stellar to halo ratios
of groups shown in Figure (\ref{fig:FG_select}),
by subdividing the groups according to 
the optical magnitude gap between the brightest and second brightest galaxy in each group,
$\Delta m_{12}$.
Each distribution is normalized to yield the integral to be unity.
}
\label{fig:stellar_mass_select}
\end{figure}


\redout{We have used, we argue, a physically more stable method to select FGs, 
as shown in Figure \ref{fig:FG_select} earlier.} \numbered{38}{We have selected FGs based on stellar mass deficiency, which we argue is less susceptible to the randomness associated with observing the $z \sim 0$ snapshot of structure formation.}
This method corresponds to a stellar to halo mass ratio of 0.005 for
groups of halo masses in the range $10^{13.4} < M_h / M_{\odot} < 10^{13.6}$,
with the remaining, stellar rich halos being the normal sample. 
Now in Figure (\ref{fig:stellar_mass_select}) 
we show the distribution of the optical magnitude gap between the brightest 
and second brightest galaxy in each group for the FGs and the normal sample.
We see that FGs tend to exhibit larger magnitude 
gaps than normal groups.
This indicates that 
this frequently used selection method based on optical magnitude gap, while not the most stable
in the context of hierarchical growth model due to random merger timing,
does show a tendency in the right sense in that the stellar deficient FGs, as a whole, 
have a larger $\Delta m_{12}$ than the normal groups.
In other words, there is indeed a broad correlation between stellar deficiency and 
$\Delta m_{12}$.
However, the difference in the stellar mass to halo mass ratio between the two samples
based on $\Delta m_{12}$ is rather small and there is a large overlap between
the two distributions.
This makes $\Delta m_{12}$ is less sensitive parameter
to untangle the stellar growth history of groups.

\begin{figure}
\plotone{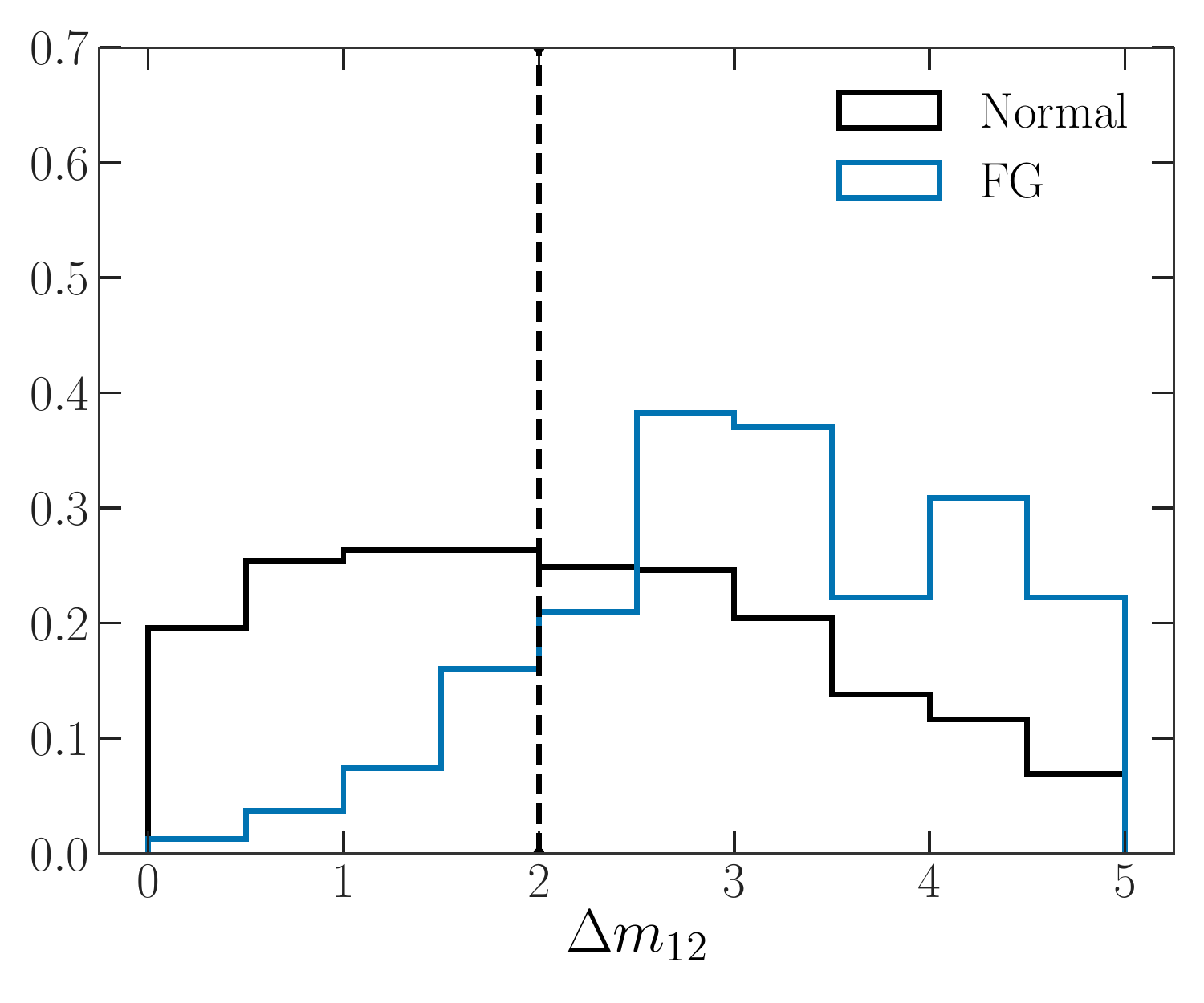}
\caption{\numbered{17 (updated figure)}{} Normalized histograms of magnitude gap in the normal and FG samples, with the traditional criteria of $\Delta m_{12}$ approximated by the dotted line. 
}
\label{fig:gap_select}
\end{figure}

Figure \ref{fig:gap_select} 
shows the distributions of $\Delta m_{12}$ for the FGs and normal groups.
Here we see that the gap $\Delta m_{12}$ for FGs
\redout{is larger and much more significant.} \numbered{41}{tends to be much larger than those of normal groups, with a typical FG selected through stellar deficiency having a magnitude gap of $\sim 3$, compared to the broad distribution peaking at $\Delta m_{12} \sim 1.5$ for gaps in the normal population.}
The comparison between Figure (\ref{fig:stellar_mass_select}) and 
Figure (\ref{fig:gap_select}) is very interesting and perhaps shows
more clearly the pros and cons of the selection methods.
\numbered{42}{The $\Delta m_{12}$ based method suggests that the groups with the largest gaps ($\Delta m_{12}>2$) tend to have ~20\% lower $M_*^{\text{group}} / M_h$, resulting in about $\sim 20\%$ of the large gap ($\Delta m_{12} > 2$) groups fulfilling our stellar deficiency criteria of $M_*^{\text{group}} / M_h > 0.005$. In contrast, our stellar deficiency 
our stellar content based selection method does produce a more pronounced
differentiation in $\Delta m_{12}$ between the two types of groups.}

Despite a simple selection method,
our results based on $\Delta m_{12}$ 
are in fact consistent with recent observations \citep{voevodkin2010, harrison2012, kundert2015},
which suggest that  optical and X-ray luminosity of $\Delta m_{12}>2$ selected fossil groups are similar to 
those of typical groups, 
since the difference in the mean $M_*^{\group}/M_h$ is a small fraction of the intrinsic scatter in the population, as we show.
It may be that historically small samples from the large magnitude gap population may have by chance exhibited a stronger difference in mass to light ratios, but larger sample sizes from more recent surveys have better probed the intrinsic scatter, resulting in two samples which are statistically consistent to within current observational limits.


\begin{figure}
\plotone{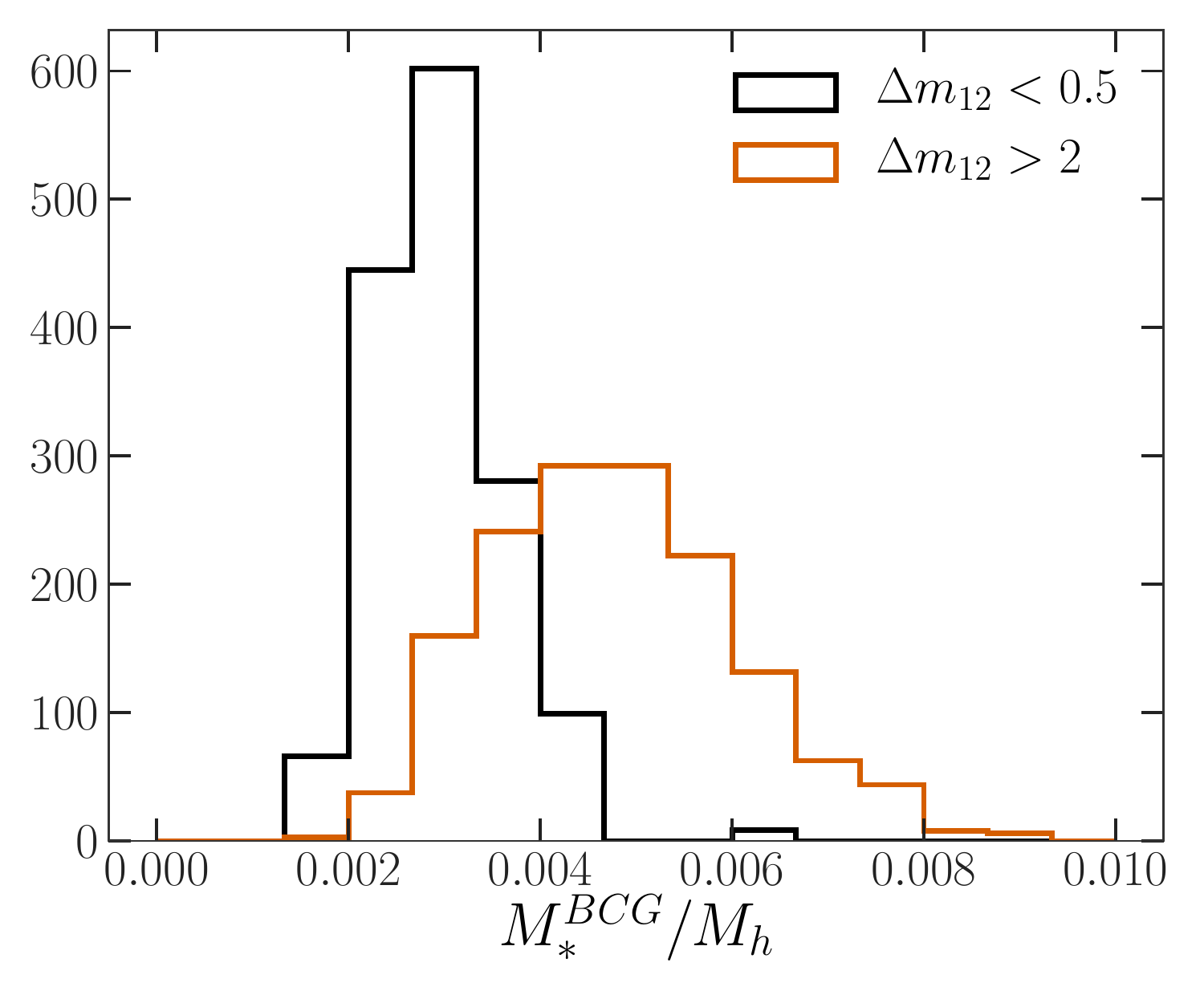}
\caption{\numbered{17}{(updated figure)}  shows the distributions of stellar mass to halo mass ratios 
based on $\Delta m_{12}$ selection.
Selecting on magnitude gap does yield a sample with a higher BCG stellar mass. \label{gap_selected_BCG}}
\end{figure}

\begin{figure}
\plotone{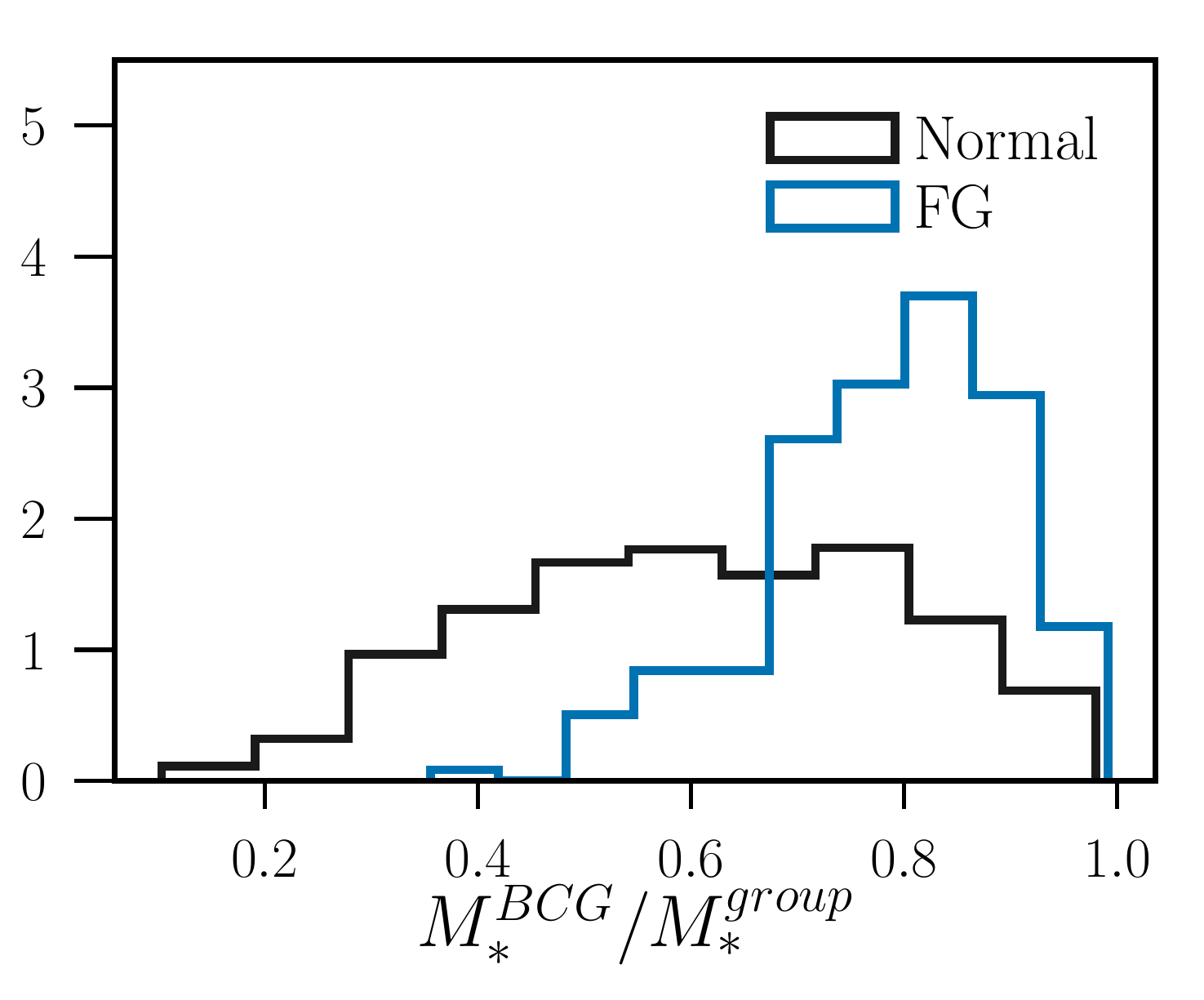}
\caption{
\numbered{46}{} The ratio of stellar mass in BCG over the total stellar mass in the groups,
for FGs (blue curve) and normal groups (black curve). These groups have been selected in the narrow halo mass range $10^{13.4} M_{\odot} < M_h < 10^{13.6}$ to control for other forms of halo bias. The union of these two populations forms the full halo population in this range.
\label{FG_select3}}
\end{figure}

\begin{figure}
\plotone{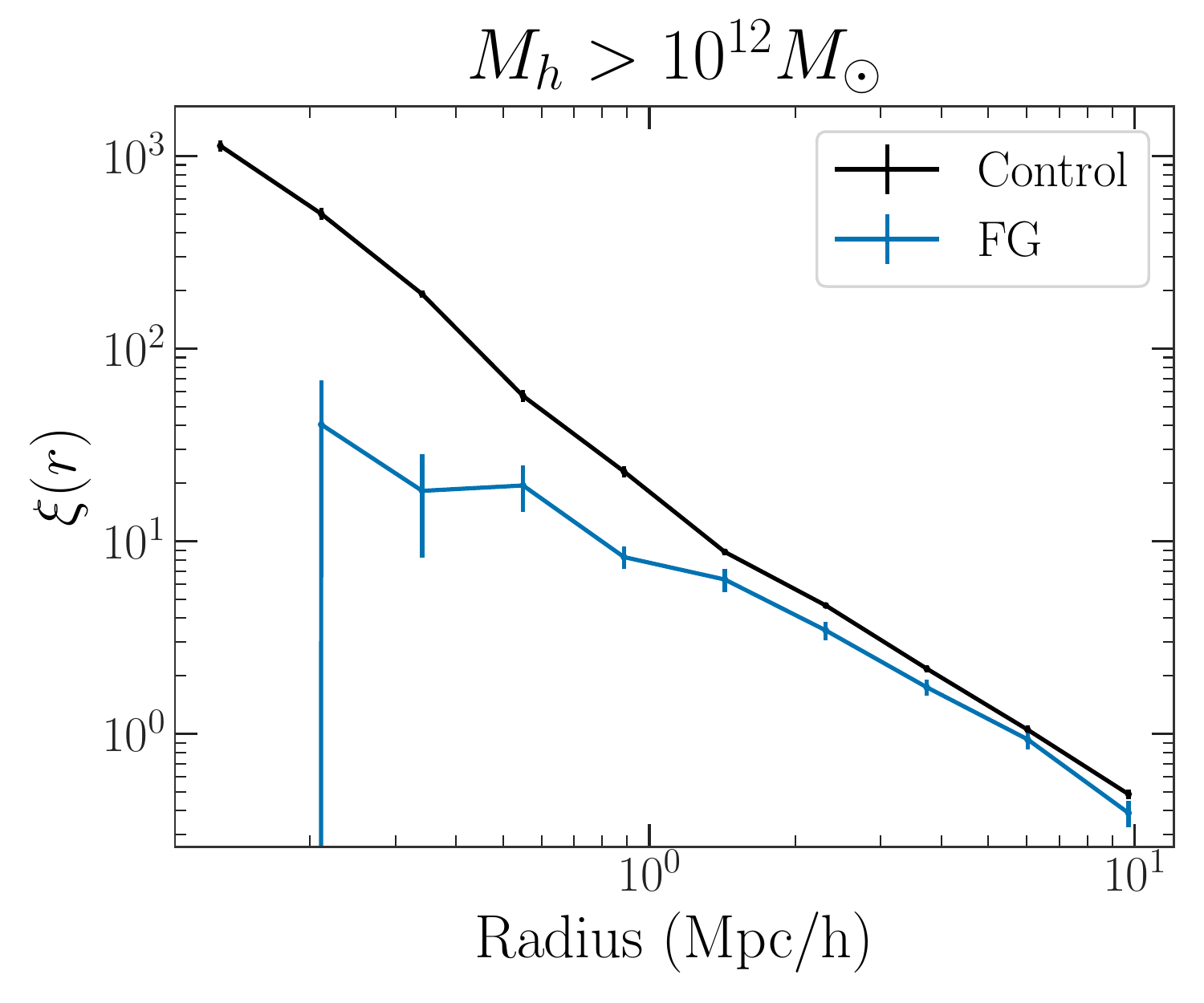}
\plotone{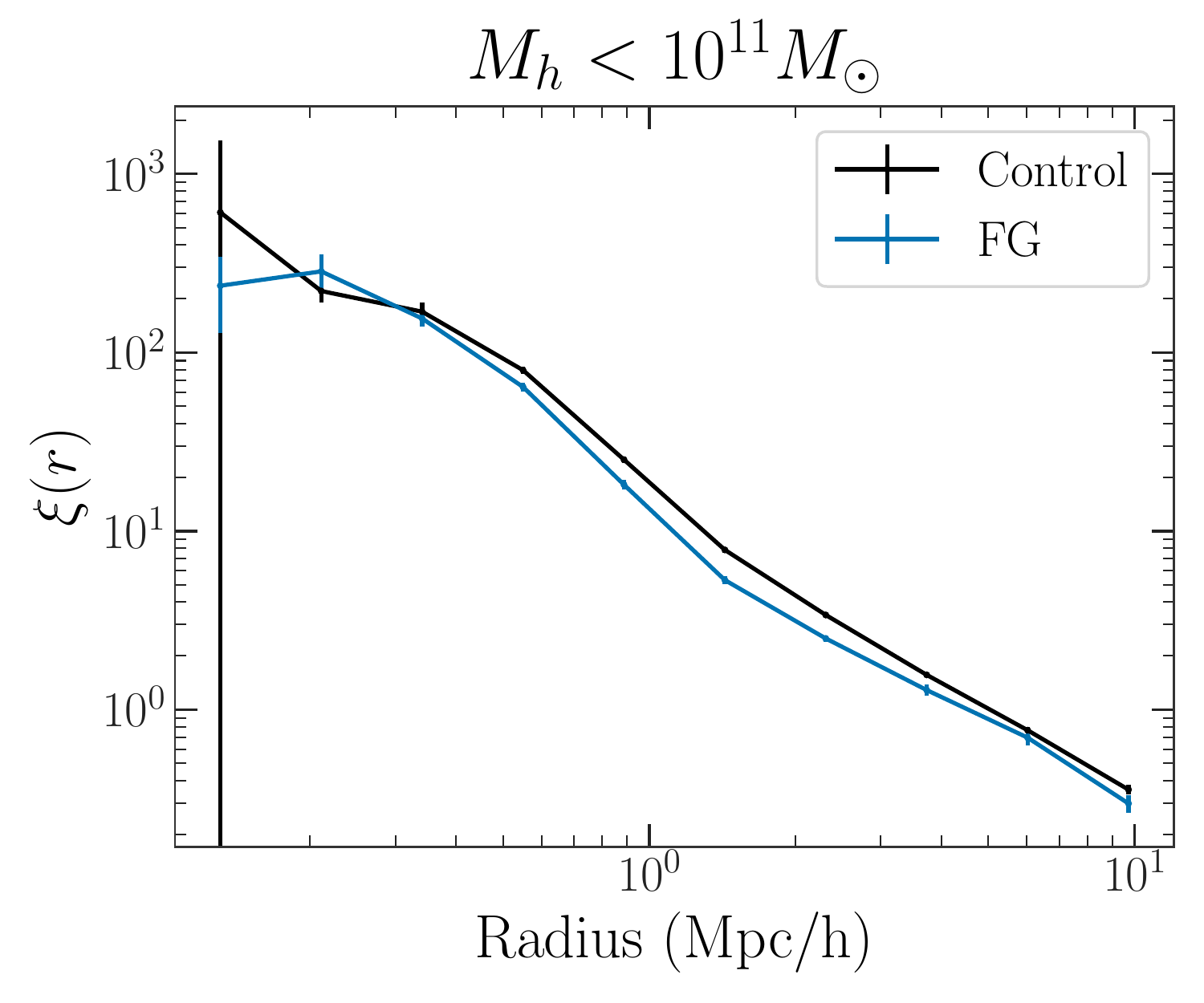}
\caption{
Top panel shows the cross correlation function between massive halos ($M_h>10^{12}\msun$)
and [FG (blue solid curve), normal groups], respectively.
FG halos tend to have fewer massive neighbors within 1 Mpc/h.
Bottom panel 
shows the cross correlation function between less massive halos ($M_h=10^{11}-10^{12}\msun$)
and [FG (blue solid curve), normal groups], respectively, where
 FGs shows only a slight deficiency in such halos within about 1Mpc/h. Error bars were performed with a jacknife using \texttt{halotools} \citep{Hearin2017}.
\label{cross_corr_massive}}
\end{figure}

\subsection{On Significant Difference in BCGs between FGs and Normal Groups} \label{BCGsection}

So far our pair-wise comparisons between FGs and normal groups
are focused on the entire group.
We now turn to the brightest central galaxies in the groups.
Given the existing literature on selecting FGs based on magnitude gap,
we first show 
in Figure \ref{gap_selected_BCG} 
the distributions of bright central galaxy stellar mass to total stellar mass ratio 
based on $\Delta m_{12}$ selection.
Interestingly, 
an opposite trend is seen here: 
groups with larger magnitude gaps tend to have brighter BCGs,
versus the trend seen in Figure (\ref{fig:stellar_mass_select}) where
groups with larger magnitude gaps tend to have lower total stellar masses.
Quite encouragingly, this result is 
consistent with observations involving the usual gap selection criteria for FGs \citep{harrison2012}, 
providing a connection between the magnitude-gap selected "fossil groups" and the groups we define as FGs. 

In Figure \ref{FG_select3} we
show the ratio of stellar mass in BCG over the total stellar mass in the groups,
for FGs (blue curve) and normal groups (black curve).
Thus, we predict that FGs have brighter BCGs compared to normal groups.
This prediction should be testable as well.

\subsection{Significant Difference in Spatial Clustering between FGs and Normal Groups} \label{sec:selection3}

One can test our model through the galaxy-group cross-correlation
between the FGs and normal group.
We show in top and bottom panels of Figure \ref{cross_corr_massive} 
the predicted cross-correlation functions between groups and two sets of satellite galaxies respectively:
relatively massive galaxies ($M_h>10^{12}\msun$) and 
less massive galaxies ($M_h<10^{11}\msun$.
We see that, 
in both cases, galaxies are less strongly clustered with FGs than
with normal groups on scales less than about 1 Mpc/h, while
on larger scales the difference is very small.
Moreover, the difference on $<1$Mpc is larger for more massive satellite galaxies 
than for less massive satellite galaxies.
Since the scale $\sim 1$Mpc is roughly about the virial radius of the groups in question,
we might summarize the above results in a more succinct way.
This is consistent with the fact that FGs form early 
that will be shown later.
More specifically, this is in excellent agreement with 
the trend seen in Figure \ref{stellarmassaccretion},
showing that, while there is continued
merging of satellites into the central BCGs of FGs,
the supply of satellites into the FGs (through the virial radii) is much lower at low redshift
compared to that of normal groups.








\section{Conclusion} \label{sec:conclusion}

We performed detailed modeling and analysis to produce a physical picture connecting the assembly history and stellar deficiency of fossil groups (FGs).
Fossil groups are typically 
We normalize our model parameters to match the observed galaxy stellar function mass function at $z=0$
and the observed star formation rate density evolution (the Madau plot).
We reach the following conclusions and predictions.

\begin{enumerate}
\item Fossil groups assemble earlier than normal groups, redshift $z\sim 2.5$ versus $z\sim 1.5$, hence the name ``fossil''.
In-situ star formation terminates earlier in fossil groups than normal groups.
Stellar mass formed in satellite halos that are ultimately accreted to reside within the virial radius of the group displays
a roughly conformal behavior to the central progenitor with respect to star formation efficiency in both cases.
Halos accreted by fossil groups form earlier than
those accreted by normal groups.
There is thus a twin-conformity between central galaxies and satellites between FGs and 
normal groups: centrals and satellites in the former form earlier and more stellar deficient
than their counterparts of the latter.
We term this effect ``assembly conformity" of halos.
This effect accounts for about 70\% of the difference in stellar content
between FGs and normal groups. \numbered{48}{This is relevant to observations focused on the magnitude gap, such as \citealt{trevisan2017}, \citealt{raouf2019}, which have suggested that large gap systems have more mergers to the central, but at higher redshift to cause no detectable imprints in SFH. }

\item
If one ``stacks" merger histories centered on the redshift 
of the peak star formation rate per unit redshift,
we find that the mass functions of satellite halos 
on either side of the peak redshift are
indistinguisable between FGs and normal groups,
indicating a self-similarity of halo assembly with respect to the peak.
Thus, the ``assembly conformity" of halos above is ultimately
a baryonic effect related to star formation efficiency but seeded by 
the timing of formation of centrals and satellies, not their halo masses.

\item
Once the central galaxies enter the hot accretion mode, it also exerts
``baryonic environmental" effects in the form of ram-pressure removal of
cold gas in satellite galaxies and heating of gas to high temperature
to prevent accretion of ample cold gas to satellite galaxies.
This effect account for about 30\% of the difference in stellar content
between FGs and normal groups.

\item In both fossil and normal groups, the formation redshift of stars in subhalos is higher than the accretion redshift, with 
the difference larger for fossil groups.
Generally, at any given halo mass, galaxies have increasingly
higher stellar to halo mass ratio with decreasing redshift.

\item
While the total stellar mass of FGs is lower than that of normal groups,
we find that the mass of the brightest central galaxy is, on average, higher than
that of normal groups. This prediction is verifiable.

\item 
We predict that in the central galaxies of FGs,
there is a negative 
stellar age gradient from the center outward,
where the opposite is expected for those in normal groups.
Moreover, the age contrast between central and satellite galaxies in FGs
is expected to be much larger than that for normal groups.
These unique predictions are important and should be testable observationally.

\item
Finally, we find that the cross-correlation function between 
Milky-size satellite halos and FGs is weaker than that 
between such satellites and normal groups by at a factor of a few 
at a separation of $0.5$Mpc (and still larger at smaller separations).
This may be observationally verifiable as well.



\end{enumerate}


\acknowledgments

\vspace{5mm}
We thank Dr. Jia Liu for kindly providing references
for the observational data used for the right panle of Figure 1.
We thank Anatoly Klypin for Bolshoi simulations that are used in our analysis.
The research is supported in part by NASA grant 80NSSC18K1101.


\clearpage

\appendix





\bibliography{bib}

\begin{thebibliography}{}
\expandafter\ifx\csname natexlab\endcsname\relax\def\natexlab#1{#1}\fi
\providecommand{\url}[1]{\href{#1}{#1}}

\bibitem[{{Adami} {et~al.}(2012){Adami}, {Jouvel}, {Guennou}, {Le Brun},
  {Durret}, {Clement}, {Clerc}, {Comer{\'o}n}, {Ilbert}, {Lin}, {Russeil}, \&
  {Seemann}}]{adami2012}
{Adami}, C., {Jouvel}, S., {Guennou}, L., {et~al.} 2012, \aap, 540, A105

\bibitem[{{Bah{\'e}} {et~al.}(2013){Bah{\'e}}, {McCarthy}, {Balogh}, \&
  {Font}}]{bahe2012}
{Bah{\'e}}, Y.~M., {McCarthy}, I.~G., {Balogh}, M.~L., \& {Font}, A.~S. 2013,
  \mnras, 430, 3017

\bibitem[{{Balogh} {et~al.}(1999){Balogh}, {Morris}, {Yee}, {Carlberg}, \&
  {Ellingson}}]{balogh1999}
{Balogh}, M.~L., {Morris}, S.~L., {Yee}, H.~K.~C., {Carlberg}, R.~G., \&
  {Ellingson}, E. 1999, \apj, 527, 54

\bibitem[{{Barnes}(1989)}]{barnes1989}
{Barnes}, J.~E. 1989, \nat, 338, 123

\bibitem[{{Benson}(2012)}]{benson2012}
{Benson}, A.~J. 2012, \na, 17, 175

\bibitem[{{Bharadwaj} {et~al.}(2016){Bharadwaj}, {Reiprich}, {Sanders}, \&
  {Schellenberger}}]{bharadwaj2016}
{Bharadwaj}, V., {Reiprich}, T.~H., {Sanders}, J.~S., \& {Schellenberger}, G.
  2016, \aap, 585, A125

\bibitem[{{Bouwens} {et~al.}(2012{\natexlab{a}}){Bouwens}, {Illingworth},
  {Oesch}, {Trenti}, {Labb{\'e}}, {Franx}, {Stiavelli}, {Carollo}, {van
  Dokkum}, \& {Magee}}]{2012Bouwens}
{Bouwens}, R.~J., {Illingworth}, G.~D., {Oesch}, P.~A., {et~al.}
  2012{\natexlab{a}}, \apjl, 752, L5

\bibitem[{{Bouwens} {et~al.}(2012{\natexlab{b}}){Bouwens}, {Illingworth},
  {Oesch}, {Franx}, {Labb{\'e}}, {Trenti}, {van Dokkum}, {Carollo},
  {Gonz{\'a}lez}, {Smit}, \& {Magee}}]{2012bBouwens}
---. 2012{\natexlab{b}}, \apj, 754, 83

\bibitem[{{Bower} {et~al.}(2006){Bower}, {Benson}, {Malbon}, {Helly}, {Frenk},
  {Baugh}, {Cole}, \& {Lacey}}]{bower2006}
{Bower}, R.~G., {Benson}, A.~J., {Malbon}, R., {et~al.} 2006, \mnras, 370, 645

\bibitem[{{Carilli} \& {Walter}(2013)}]{2013Carilli}
{Carilli}, C.~L., \& {Walter}, F. 2013, \araa, 51, 105

\bibitem[{{Cen}(2014)}]{cen2014}
{Cen}, R. 2014, \apj, 789, L21

\bibitem[{{Crain} {et~al.}(2009){Crain}, {Theuns}, {Dalla Vecchia}, {Eke},
  {Frenk}, {Jenkins}, {Kay}, {Peacock}, {Pearce}, {Schaye}, {Springel},
  {Thomas}, {White}, \& {Wiersma}}]{crain2009}
{Crain}, R.~A., {Theuns}, T., {Dalla Vecchia}, C., {et~al.} 2009, \mnras, 399,
  1773

\bibitem[{{Cucciati} {et~al.}(2012){Cucciati}, {Tresse}, {Ilbert}, {Le
  F{\`e}vre}, {Garilli}, {Le Brun}, {Cassata}, {Franzetti}, {Maccagni},
  {Scodeggio}, {Zucca}, {Zamorani}, {Bardelli}, {Bolzonella}, {Bielby},
  {McCracken}, {Zanichelli}, \& {Vergani}}]{2012Cucciati}
{Cucciati}, O., {Tresse}, L., {Ilbert}, O., {et~al.} 2012, \aap, 539, A31

\bibitem[{{Cui} {et~al.}(2011){Cui}, {Springel}, {Yang}, {De Lucia}, \&
  {Borgani}}]{cui2011}
{Cui}, W., {Springel}, V., {Yang}, X., {De Lucia}, G., \& {Borgani}, S. 2011,
  \mnras, 416, 2997

\bibitem[{{Dahlen} {et~al.}(2007){Dahlen}, {Mobasher}, {Dickinson}, {Ferguson},
  {Giavalisco}, {Kretchmer}, \& {Ravindranath}}]{2007Dahlen}
{Dahlen}, T., {Mobasher}, B., {Dickinson}, M., {et~al.} 2007, \apj, 654, 172

\bibitem[{{Dariush} {et~al.}(2007){Dariush}, {Khosroshahi}, {Ponman}, {Pearce},
  {Raychaudhury}, \& {Hartley}}]{dariush2007}
{Dariush}, A., {Khosroshahi}, H.~G., {Ponman}, T.~J., {et~al.} 2007, \mnras,
  382, 433

\bibitem[{{Dariush} {et~al.}(2010){Dariush}, {Raychaudhury}, {Ponman},
  {Khosroshahi}, {Benson}, {Bower}, \& {Pearce}}]{dariush2010}
{Dariush}, A.~A., {Raychaudhury}, S., {Ponman}, T.~J., {et~al.} 2010, \mnras,
  405, 1873

\bibitem[{{D{\'\i}az-Gim{\'e}nez} {et~al.}(2011){D{\'\i}az-Gim{\'e}nez},
  {Zandivarez}, {Proctor}, {Mendes de Oliveira}, \& {Abramo}}]{diazgimenez2011}
{D{\'\i}az-Gim{\'e}nez}, E., {Zandivarez}, A., {Proctor}, R., {Mendes de
  Oliveira}, C., \& {Abramo}, L.~R. 2011, \aap, 527, A129

\bibitem[{{D'Onghia} {et~al.}(2005){D'Onghia}, {Sommer-Larsen}, {Romeo},
  {Burkert}, {Pedersen}, {Portinari}, \& {Rasmussen}}]{donghia2005}
{D'Onghia}, E., {Sommer-Larsen}, J., {Romeo}, A.~D., {et~al.} 2005, \apj, 630,
  L109

\bibitem[{{Dunne} {et~al.}(2009){Dunne}, {Ivison}, {Maddox}, {Cirasuolo},
  {Mortier}, {Foucaud}, {Ibar}, {Almaini}, {Simpson}, \& {McLure}}]{2009Dunne}
{Dunne}, L., {Ivison}, R.~J., {Maddox}, S., {et~al.} 2009, \mnras, 394, 3

\bibitem[{{Eigenthaler} \& {Zeilinger}(2013)}]{eigenthaler2013}
{Eigenthaler}, P., \& {Zeilinger}, W.~W. 2013, \aap, 553, A99

\bibitem[{{Gozaliasl} {et~al.}(2014){Gozaliasl}, {Khosroshahi}, {Dariush},
  {Finoguenov}, {Jassur}, \& {Molaeinezhad}}]{gozaliasl2014}
{Gozaliasl}, G., {Khosroshahi}, H.~G., {Dariush}, A.~A., {et~al.} 2014, \aap,
  571, A49

\bibitem[{{Gruppioni} {et~al.}(2013){Gruppioni}, {Pozzi}, {Rodighiero},
  {Delvecchio}, {Berta}, {Pozzetti}, {Zamorani}, {Andreani}, {Cimatti},
  {Ilbert}, {Le Floc'h}, {Lutz}, {Magnelli}, {Marchetti}, {Monaco}, {Nordon},
  {Oliver}, {Popesso}, {Riguccini}, {Roseboom}, {Rosario}, {Sargent},
  {Vaccari}, {Altieri}, {Aussel}, {Bongiovanni}, {Cepa}, {Daddi},
  {Dom{\'{\i}}nguez-S{\'a}nchez}, {Elbaz}, {F{\"o}rster Schreiber}, {Genzel},
  {Iribarrem}, {Magliocchetti}, {Maiolino}, {Poglitsch}, {P{\'e}rez
  Garc{\'{\i}}a}, {Sanchez-Portal}, {Sturm}, {Tacconi}, {Valtchanov},
  {Amblard}, {Arumugam}, {Bethermin}, {Bock}, {Boselli}, {Buat}, {Burgarella},
  {Castro-Rodr{\'{\i}}guez}, {Cava}, {Chanial}, {Clements}, {Conley}, {Cooray},
  {Dowell}, {Dwek}, {Eales}, {Franceschini}, {Glenn}, {Griffin},
  {Hatziminaoglou}, {Ibar}, {Isaak}, {Ivison}, {Lagache}, {Levenson}, {Lu},
  {Madden}, {Maffei}, {Mainetti}, {Nguyen}, {O'Halloran}, {Page}, {Panuzzo},
  {Papageorgiou}, {Pearson}, {P{\'e}rez-Fournon}, {Pohlen}, {Rigopoulou},
  {Rowan-Robinson}, {Schulz}, {Scott}, {Seymour}, {Shupe}, {Smith}, {Stevens},
  {Symeonidis}, {Trichas}, {Tugwell}, {Vigroux}, {Wang}, {Wright}, {Xu},
  {Zemcov}, {Bardelli}, {Carollo}, {Contini}, {Le F{\'e}vre}, {Lilly},
  {Mainieri}, {Renzini}, {Scodeggio}, \& {Zucca}}]{2013Gruppioni}
{Gruppioni}, C., {Pozzi}, F., {Rodighiero}, G., {et~al.} 2013, \mnras, 432, 23

\bibitem[{{Haines} {et~al.}(2015){Haines}, {Pereira}, {Smith}, {Egami},
  {Babul}, {Finoguenov}, {Ziparo}, {McGee}, {Rawle}, {Okabe}, \&
  {Moran}}]{haines2015}
{Haines}, C.~P., {Pereira}, M.~J., {Smith}, G.~P., {et~al.} 2015, \apj, 806,
  101

\bibitem[{{Harrison} {et~al.}(2012){Harrison}, {Miller}, {Richards},
  {Lloyd-Davies}, {Hoyle}, {Romer}, {Mehrtens}, {Hilton}, {Stott}, {Capozzi},
  {Collins}, {Deadman}, {Liddle}, {Sahl{\'e}n}, {Stanford}, \&
  {Viana}}]{harrison2012}
{Harrison}, C.~D., {Miller}, C.~J., {Richards}, J.~W., {et~al.} 2012, \apj,
  752, 12

\bibitem[{{Hearin} {et~al.}(2016){Hearin}, {Behroozi}, \& {van den
  Bosch}}]{hearin2016}
{Hearin}, A.~P., {Behroozi}, P.~S., \& {van den Bosch}, F.~C. 2016, \mnras,
  461, 2135

\bibitem[{{Hearin} {et~al.}(2017){Hearin}, {Campbell}, {Tollerud}, {Behroozi},
  {Diemer}, {Goldbaum}, {Jennings}, {Leauthaud}, {Mao}, {More}, {Parejko},
  {Sinha}, {Sip{\"o}cz}, \& {Zentner}}]{Hearin2017}
{Hearin}, A.~P., {Campbell}, D., {Tollerud}, E., {et~al.} 2017, \aj, 154, 190

\bibitem[{{Henriques} {et~al.}(2015){Henriques}, {White}, {Thomas}, {Angulo},
  {Guo}, {Lemson}, {Springel}, \& {Overzier}}]{henriques2015}
{Henriques}, B. M.~B., {White}, S. D.~M., {Thomas}, P.~A., {et~al.} 2015,
  \mnras, 451, 2663

\bibitem[{{Jones} {et~al.}(2003){Jones}, {Ponman}, {Horton}, {Babul},
  {Ebeling}, \& {Burke}}]{jones2003}
{Jones}, L.~R., {Ponman}, T.~J., {Horton}, A., {et~al.} 2003, \mnras, 343, 627

\bibitem[{{Kajisawa} {et~al.}(2010){Kajisawa}, {Ichikawa}, {Yamada},
  {Uchimoto}, {Yoshikawa}, {Akiyama}, \& {Onodera}}]{2010Kajisawa}
{Kajisawa}, M., {Ichikawa}, T., {Yamada}, T., {et~al.} 2010, \apj, 723, 129

\bibitem[{{Kanagusuku} {et~al.}(2016){Kanagusuku}, {D{\'\i}az-Gim{\'e}nez}, \&
  {Zandivarez}}]{kanagusuku2016}
{Kanagusuku}, M.~J., {D{\'\i}az-Gim{\'e}nez}, E., \& {Zandivarez}, A. 2016,
  \aap, 586, A40

\bibitem[{{Karim} {et~al.}(2011){Karim}, {Schinnerer},
  {Mart{\'{\i}}nez-Sansigre}, {Sargent}, {van der Wel}, {Rix}, {Ilbert},
  {Smol{\v c}i{\'c}}, {Carilli}, {Pannella}, {Koekemoer}, {Bell}, \&
  {Salvato}}]{2011Karim}
{Karim}, A., {Schinnerer}, E., {Mart{\'{\i}}nez-Sansigre}, A., {et~al.} 2011,
  \apj, 730, 61

\bibitem[{{Katz}(1992)}]{katz1992}
{Katz}, N. 1992, \apj, 391, 502

\bibitem[{{Kauffmann} {et~al.}(1993){Kauffmann}, {White}, \&
  {Guiderdoni}}]{kauffmann1993}
{Kauffmann}, G., {White}, S.~D.~M., \& {Guiderdoni}, B. 1993, \mnras, 264, 201

\bibitem[{{Kere{\v{s}}} {et~al.}(2009){Kere{\v{s}}}, {Katz}, {Fardal},
  {Dav{\'e}}, \& {Weinberg}}]{keres2009}
{Kere{\v{s}}}, D., {Katz}, N., {Fardal}, M., {Dav{\'e}}, R., \& {Weinberg},
  D.~H. 2009, \mnras, 395, 160

\bibitem[{{Khosroshahi} {et~al.}(2014){Khosroshahi}, {Gozaliasl}, {Rasmussen},
  {Molaeinezhad}, {Ponman}, {Dariush}, \& {Sanderson}}]{khosroshahi2014}
{Khosroshahi}, H.~G., {Gozaliasl}, G., {Rasmussen}, J., {et~al.} 2014, \mnras,
  443, 318

\bibitem[{{Khosroshahi} {et~al.}(2007){Khosroshahi}, {Ponman}, \&
  {Jones}}]{khosroshahi2007}
{Khosroshahi}, H.~G., {Ponman}, T.~J., \& {Jones}, L.~R. 2007, \mnras, 377, 595

\bibitem[{{Khosroshahi} {et~al.}(2017){Khosroshahi}, {Raouf}, {Miraghaei},
  {Brough}, {Croton}, {Driver}, {Graham}, {Baldry}, {Brown}, {Prescott}, \&
  {Wang}}]{khosroshahi2017}
{Khosroshahi}, H.~G., {Raouf}, M., {Miraghaei}, H., {et~al.} 2017, \apj, 842,
  81

\bibitem[{{Klypin} {et~al.}(2016){Klypin}, {Yepes}, {Gottl{\"o}ber}, {Prada},
  \& {He{\ss}}}]{klypin2016}
{Klypin}, A., {Yepes}, G., {Gottl{\"o}ber}, S., {Prada}, F., \& {He{\ss}}, S.
  2016, \mnras, 457, 4340

\bibitem[{{Klypin} {et~al.}(2011){Klypin}, {Trujillo-Gomez}, \&
  {Primack}}]{klypin2011}
{Klypin}, A.~A., {Trujillo-Gomez}, S., \& {Primack}, J. 2011, \apj, 740, 102

\bibitem[{{Krick} \& {Bernstein}(2007)}]{krick2007}
{Krick}, J.~E., \& {Bernstein}, R.~A. 2007, \aj, 134, 466

\bibitem[{{Kundert} {et~al.}(2017){Kundert}, {D'Onghia}, \&
  {Aguerri}}]{kundert2017}
{Kundert}, A., {D'Onghia}, E., \& {Aguerri}, J.~A.~L. 2017, \apj, 845, 45

\bibitem[{{Kundert} {et~al.}(2015){Kundert}, {Gastaldello}, {D'Onghia},
  {Girardi}, {Aguerri}, {Barrena}, {Corsini}, {De Grandi},
  {Jim{\'e}nez-Bail{\'o}n}, {Lozada-Mu{\~n}oz}, {M{\'e}ndez-Abreu}, {S{\'a
  }nchez-Janssen}, {Wilcots}, \& {Zarattini}}]{kundert2015}
{Kundert}, A., {Gastaldello}, F., {D'Onghia}, E., {et~al.} 2015, \mnras, 454,
  161

\bibitem[{{La Barbera} {et~al.}(2009){La Barbera}, {de Carvalho}, {de la Rosa},
  {Sorrentino}, {Gal}, \& {Kohl-Moreira}}]{labarbera2009}
{La Barbera}, F., {de Carvalho}, R.~R., {de la Rosa}, I.~G., {et~al.} 2009,
  \aj, 137, 3942

\bibitem[{{Le Borgne} {et~al.}(2009){Le Borgne}, {Elbaz}, {Ocvirk}, \&
  {Pichon}}]{2009LeBorgne}
{Le Borgne}, D., {Elbaz}, D., {Ocvirk}, P., \& {Pichon}, C. 2009, \aap, 504,
  727

\bibitem[{{Ly} {et~al.}(2011{\natexlab{a}}){Ly}, {Lee}, {Dale}, {Momcheva},
  {Salim}, {Staudaher}, {Moore}, \& {Finn}}]{2011Ly}
{Ly}, C., {Lee}, J.~C., {Dale}, D.~A., {et~al.} 2011{\natexlab{a}}, \apj, 726,
  109

\bibitem[{{Ly} {et~al.}(2011{\natexlab{b}}){Ly}, {Malkan}, {Hayashi},
  {Motohara}, {Kashikawa}, {Shimasaku}, {Nagao}, \& {Grady}}]{2011bLy}
{Ly}, C., {Malkan}, M.~A., {Hayashi}, M., {et~al.} 2011{\natexlab{b}}, \apj,
  735, 91

\bibitem[{{Magnelli} {et~al.}(2011){Magnelli}, {Elbaz}, {Chary}, {Dickinson},
  {Le Borgne}, {Frayer}, \& {Willmer}}]{2011Magnelli}
{Magnelli}, B., {Elbaz}, D., {Chary}, R.~R., {et~al.} 2011, \aap, 528, A35

\bibitem[{{Magnelli} {et~al.}(2013){Magnelli}, {Popesso}, {Berta}, {Pozzi},
  {Elbaz}, {Lutz}, {Dickinson}, {Altieri}, {Andreani}, {Aussel},
  {B{\'e}thermin}, {Bongiovanni}, {Cepa}, {Charmandaris}, {Chary}, {Cimatti},
  {Daddi}, {F{\"o}rster Schreiber}, {Genzel}, {Gruppioni}, {Harwit}, {Hwang},
  {Ivison}, {Magdis}, {Maiolino}, {Murphy}, {Nordon}, {Pannella}, {P{\'e}rez
  Garc{\'{\i}}a}, {Poglitsch}, {Rosario}, {Sanchez-Portal}, {Santini}, {Scott},
  {Sturm}, {Tacconi}, \& {Valtchanov}}]{2013Magnelli}
{Magnelli}, B., {Popesso}, P., {Berta}, S., {et~al.} 2013, \aap, 553, A132

\bibitem[{{Miller} {et~al.}(2012){Miller}, {Rykoff}, {Dupke}, {Mendes de
  Oliveira}, {Lopes de Oliveira}, {Proctor}, {Garmire}, {Koester}, \&
  {McKay}}]{miller2012}
{Miller}, E.~D., {Rykoff}, E.~S., {Dupke}, R.~A., {et~al.} 2012, \apj, 747, 94

\bibitem[{{Moustakas} {et~al.}(2013){Moustakas}, {Coil}, {Aird}, {Blanton},
  {Cool}, {Eisenstein}, {Mendez}, {Wong}, {Zhu}, \& {Arnouts}}]{moustakas2013}
{Moustakas}, J., {Coil}, A.~L., {Aird}, J., {et~al.} 2013, \apj, 767, 50

\bibitem[{{Pierini} {et~al.}(2011){Pierini}, {Giodini}, {Finoguenov},
  {B{\"o}hringer}, {D'Onghia}, {Pratt}, {D{\'e}mocl{\`e}s}, {Pannella},
  {Zibetti}, {Braglia}, {Verdugo}, {Ziparo}, {Koekemoer}, {Salvato}, \& {COSMOS
  Collaboration}}]{pierini2011}
{Pierini}, D., {Giodini}, S., {Finoguenov}, A., {et~al.} 2011, \mnras, 417,
  2927

\bibitem[{{Pillepich} {et~al.}(2018){Pillepich}, {Springel}, {Nelson}, {Genel},
  {Naiman}, {Pakmor}, {Hernquist}, {Torrey}, {Vogelsberger}, {Weinberger}, \&
  {Marinacci}}]{pillepich2018}
{Pillepich}, A., {Springel}, V., {Nelson}, D., {et~al.} 2018, \mnras, 473, 4077

\bibitem[{{Planck Collaboration} {et~al.}(2014){Planck Collaboration}, {Ade},
  {Aghanim}, {Armitage-Caplan}, {Arnaud}, {Ashdown}, {Atrio-Barandela},
  {Aumont}, {Baccigalupi}, {Banday}, \& et~al.}]{2014PlanckParams}
{Planck Collaboration}, {Ade}, P.~A.~R., {Aghanim}, N., {et~al.} 2014, \aap,
  571, A16

\bibitem[{{Proctor} {et~al.}(2011){Proctor}, {de Oliveira}, {Dupke}, {de
  Oliveira}, {Cypriano}, {Miller}, \& {Rykoff}}]{proctor2011}
{Proctor}, R.~N., {de Oliveira}, C.~M., {Dupke}, R., {et~al.} 2011, \mnras,
  418, 2054

\bibitem[{{Proctor} {et~al.}(2014){Proctor}, {Mendes de Oliveira}, \&
  {Eigenthaler}}]{proctor2014}
{Proctor}, R.~N., {Mendes de Oliveira}, C., \& {Eigenthaler}, P. 2014, \mnras,
  439, 2281

\bibitem[{{Raouf} {et~al.}(2018){Raouf}, {Khosroshahi}, {Mamon}, {Croton},
  {Hashemizadeh}, \& {Dariush}}]{raouf2018}
{Raouf}, M., {Khosroshahi}, H.~G., {Mamon}, G.~A., {et~al.} 2018, \apj, 863, 40

\bibitem[{{Raouf} {et~al.}(2014){Raouf}, {Khosroshahi}, {Ponman}, {Dariush},
  {Molaeinezhad}, \& {Tavasoli}}]{raouf2014}
{Raouf}, M., {Khosroshahi}, H.~G., {Ponman}, T.~J., {et~al.} 2014, \mnras, 442,
  1578

\bibitem[{{Raouf} {et~al.}(2017){Raouf}, {Shabala}, {Croton}, {Khosroshahi}, \&
  {Bernyk}}]{raouf2017}
{Raouf}, M., {Shabala}, S.~S., {Croton}, D.~J., {Khosroshahi}, H.~G., \&
  {Bernyk}, M. 2017, \mnras, 471, 658

\bibitem[{{Raouf} {et~al.}(2019){Raouf}, {Smith}, {Khosroshahi}, {Dariush},
  {Driver}, {Ko}, \& {Hwang}}]{raouf2019}
{Raouf}, M., {Smith}, R., {Khosroshahi}, H.~G., {et~al.} 2019, \apj, 887, 264

\bibitem[{{Reddy} \& {Steidel}(2009)}]{2009Reddy}
{Reddy}, N.~A., \& {Steidel}, C.~C. 2009, \apj, 692, 778

\bibitem[{{Robotham} \& {Driver}(2011)}]{2011Robotham}
{Robotham}, A.~S.~G., \& {Driver}, S.~P. 2011, \mnras, 413, 2570

\bibitem[{{Rujopakarn} {et~al.}(2010){Rujopakarn}, {Eisenstein}, {Rieke},
  {Papovich}, {Cool}, {Moustakas}, {Jannuzi}, {Kochanek}, {Rieke}, {Dey},
  {Eisenhardt}, {Murray}, {Brown}, \& {Le Floc'h}}]{2010Rujopakarn}
{Rujopakarn}, W., {Eisenstein}, D.~J., {Rieke}, G.~H., {et~al.} 2010, \apj,
  718, 1171

\bibitem[{{Salim} {et~al.}(2007){Salim}, {Rich}, {Charlot}, {Brinchmann},
  {Johnson}, {Schiminovich}, {Seibert}, {Mallery}, {Heckman}, {Forster},
  {Friedman}, {Martin}, {Morrissey}, {Neff}, {Small}, {Wyder}, {Bianchi},
  {Donas}, {Lee}, {Madore}, {Milliard}, {Szalay}, {Welsh}, \& {Yi}}]{2007Salim}
{Salim}, S., {Rich}, R.~M., {Charlot}, S., {et~al.} 2007, \apjs, 173, 267

\bibitem[{{Sanders} {et~al.}(2003){Sanders}, {Mazzarella}, {Kim}, {Surace}, \&
  {Soifer}}]{2003Sanders}
{Sanders}, D.~B., {Mazzarella}, J.~M., {Kim}, D.-C., {Surace}, J.~A., \&
  {Soifer}, B.~T. 2003, \aj, 126, 1607

\bibitem[{{Schaye} {et~al.}(2010){Schaye}, {Dalla Vecchia}, {Booth}, {Wiersma},
  {Theuns}, {Haas}, {Bertone}, {Duffy}, {McCarthy}, \& {van de
  Voort}}]{schaye2010}
{Schaye}, J., {Dalla Vecchia}, C., {Booth}, C.~M., {et~al.} 2010, \mnras, 402,
  1536

\bibitem[{{Schenker} {et~al.}(2013){Schenker}, {Robertson}, {Ellis}, {Ono},
  {McLure}, {Dunlop}, {Koekemoer}, {Bowler}, {Ouchi}, {Curtis-Lake}, {Rogers},
  {Schneider}, {Charlot}, {Stark}, {Furlanetto}, \& {Cirasuolo}}]{2013Schenker}
{Schenker}, M.~A., {Robertson}, B.~E., {Ellis}, R.~S., {et~al.} 2013, \apj,
  768, 196

\bibitem[{{Schiminovich} {et~al.}(2005){Schiminovich}, {Ilbert}, {Arnouts},
  {Milliard}, {Tresse}, {Le F{\`e}vre}, {Treyer}, {Wyder}, {Budav{\'a}ri},
  {Zucca}, {Zamorani}, {Martin}, {Adami}, {Arnaboldi}, {Bardelli}, {Barlow},
  {Bianchi}, {Bolzonella}, {Bottini}, {Byun}, {Cappi}, {Contini}, {Charlot},
  {Donas}, {Forster}, {Foucaud}, {Franzetti}, {Friedman}, {Garilli},
  {Gavignaud}, {Guzzo}, {Heckman}, {Hoopes}, {Iovino}, {Jelinsky}, {Le Brun},
  {Lee}, {Maccagni}, {Madore}, {Malina}, {Marano}, {Marinoni}, {McCracken},
  {Mazure}, {Meneux}, {Morrissey}, {Neff}, {Paltani}, {Pell{\`o}}, {Picat},
  {Pollo}, {Pozzetti}, {Radovich}, {Rich}, {Scaramella}, {Scodeggio},
  {Seibert}, {Siegmund}, {Small}, {Szalay}, {Vettolani}, {Welsh}, {Xu}, \&
  {Zanichelli}}]{2005Schiminovich}
{Schiminovich}, D., {Ilbert}, O., {Arnouts}, S., {et~al.} 2005, \apjl, 619, L47

\bibitem[{{Shim} {et~al.}(2009){Shim}, {Colbert}, {Teplitz}, {Henry}, {Malkan},
  {McCarthy}, \& {Yan}}]{2009Shim}
{Shim}, H., {Colbert}, J., {Teplitz}, H., {et~al.} 2009, \apj, 696, 785

\bibitem[{{Smol{\v c}i{\'c}} {et~al.}(2009){Smol{\v c}i{\'c}}, {Schinnerer},
  {Zamorani}, {Bell}, {Bondi}, {Carilli}, {Ciliegi}, {Mobasher}, {Paglione},
  {Scodeggio}, \& {Scoville}}]{2009Smolvic}
{Smol{\v c}i{\'c}}, V., {Schinnerer}, E., {Zamorani}, G., {et~al.} 2009, \apj,
  690, 610

\bibitem[{{Sobral} {et~al.}(2013){Sobral}, {Smail}, {Best}, {Geach}, {Matsuda},
  {Stott}, {Cirasuolo}, \& {Kurk}}]{2013Sobral}
{Sobral}, D., {Smail}, I., {Best}, P.~N., {et~al.} 2013, \mnras, 428, 1128

\bibitem[{{Somerville} \& {Primack}(1999)}]{somerville1999}
{Somerville}, R.~S., \& {Primack}, J.~R. 1999, \mnras, 310, 1087

\bibitem[{{Springel}(2005)}]{springel2005}
{Springel}, V. 2005, \mnras, 364, 1105

\bibitem[{{Tadaki} {et~al.}(2011){Tadaki}, {Kodama}, {Koyama}, {Hayashi},
  {Tanaka}, \& {Tokoku}}]{2011Tadaki}
{Tadaki}, K.-I., {Kodama}, T., {Koyama}, Y., {et~al.} 2011, \pasj, 63, 437

\bibitem[{{Takeuchi} {et~al.}(2003){Takeuchi}, {Yoshikawa}, \&
  {Ishii}}]{2003Takeuchi}
{Takeuchi}, T.~T., {Yoshikawa}, K., \& {Ishii}, T.~T. 2003, \apjl, 587, L89

\bibitem[{{Trevisan} {et~al.}(2017){Trevisan}, {Mamon}, \&
  {Khosroshahi}}]{trevisan2017}
{Trevisan}, M., {Mamon}, G.~A., \& {Khosroshahi}, H.~G. 2017, \mnras, 464, 4593

\bibitem[{{Trujillo-Gomez} {et~al.}(2011){Trujillo-Gomez}, {Klypin}, {Primack},
  \& {Romanowsky}}]{trujillogomez2011}
{Trujillo-Gomez}, S., {Klypin}, A., {Primack}, J., \& {Romanowsky}, A.~J. 2011,
  \apj, 742, 16

\bibitem[{{van der Burg} {et~al.}(2010){van der Burg}, {Hildebrandt}, \&
  {Erben}}]{2010vanderBurg}
{van der Burg}, R.~F.~J., {Hildebrandt}, H., \& {Erben}, T. 2010, \aap, 523,
  A74

\bibitem[{{Vikhlinin} {et~al.}(1999){Vikhlinin}, {McNamara}, {Hornstrup},
  {Quintana}, {Forman}, {Jones}, \& {Way}}]{vikhlinin1999}
{Vikhlinin}, A., {McNamara}, B.~R., {Hornstrup}, A., {et~al.} 1999, \apj, 520,
  L1

\bibitem[{{Voevodkin} {et~al.}(2010){Voevodkin}, {Borozdin}, {Heitmann},
  {Habib}, {Vikhlinin}, {Mescheryakov}, {Hornstrup}, \&
  {Burenin}}]{voevodkin2010}
{Voevodkin}, A., {Borozdin}, K., {Heitmann}, K., {et~al.} 2010, \apj, 708, 1376

\bibitem[{{Vogelsberger} {et~al.}(2020){Vogelsberger}, {Marinacci}, {Torrey},
  \& {Puchwein}}]{vogelsberger2020}
{Vogelsberger}, M., {Marinacci}, F., {Torrey}, P., \& {Puchwein}, E. 2020,
  Nature Reviews Physics, 2, 42

\bibitem[{{Vogelsberger} {et~al.}(2014){Vogelsberger}, {Genel}, {Springel},
  {Torrey}, {Sijacki}, {Xu}, {Snyder}, {Nelson}, \&
  {Hernquist}}]{vogelsberger2014}
{Vogelsberger}, M., {Genel}, S., {Springel}, V., {et~al.} 2014, \mnras, 444,
  1518

\bibitem[{{von Benda-Beckmann} {et~al.}(2008){von Benda-Beckmann}, {D'Onghia},
  {Gottl{\"o}ber}, {Hoeft}, {Khalatyan}, {Klypin}, \&
  {M{\"u}ller}}]{vonbendabeckmann2008}
{von Benda-Beckmann}, A.~M., {D'Onghia}, E., {Gottl{\"o}ber}, S., {et~al.}
  2008, \mnras, 386, 2345

\bibitem[{{Wetzel} {et~al.}(2012){Wetzel}, {Tinker}, \& {Conroy}}]{wetzel2012}
{Wetzel}, A.~R., {Tinker}, J.~L., \& {Conroy}, C. 2012, \mnras, 424, 232

\bibitem[{{Wyder} {et~al.}(2005){Wyder}, {Treyer}, {Milliard}, {Schiminovich},
  {Arnouts}, {Budav{\'a}ri}, {Barlow}, {Bianchi}, {Byun}, {Donas}, {Forster},
  {Friedman}, {Heckman}, {Jelinsky}, {Lee}, {Madore}, {Malina}, {Martin},
  {Morrissey}, {Neff}, {Rich}, {Siegmund}, {Small}, {Szalay}, \&
  {Welsh}}]{2005Wyder}
{Wyder}, T.~K., {Treyer}, M.~A., {Milliard}, B., {et~al.} 2005, \apjl, 619, L15

\bibitem[{{Yoshida} {et~al.}(2006){Yoshida}, {Shimasaku}, {Kashikawa}, {Ouchi},
  {Okamura}, {Ajiki}, {Akiyama}, {Ando}, {Aoki}, {Doi}, {Furusawa},
  {Hayashino}, {Iwamuro}, {Iye}, {Karoji}, {Kobayashi}, {Kodaira}, {Kodama},
  {Komiyama}, {Malkan}, {Matsuda}, {Miyazaki}, {Mizumoto}, {Morokuma},
  {Motohara}, {Murayama}, {Nagao}, {Nariai}, {Ohta}, {Sasaki}, {Sato},
  {Sekiguchi}, {Shioya}, {Tamura}, {Taniguchi}, {Umemura}, {Yamada}, \&
  {Yasuda}}]{2006Yoshida}
{Yoshida}, M., {Shimasaku}, K., {Kashikawa}, N., {et~al.} 2006, \apj, 653, 988

\bibitem[{{Yoshioka} {et~al.}(2004){Yoshioka}, {Furuzawa}, {Takahashi},
  {Tawara}, {Sato}, {Yamashita}, \& {Kumai}}]{yoshioka2004}
{Yoshioka}, T., {Furuzawa}, A., {Takahashi}, S., {et~al.} 2004, Advances in
  Space Research, 34, 2525

\bibitem[{{Zarattini} {et~al.}(2016){Zarattini}, {Girardi}, {Aguerri},
  {Boschin}, {Barrena}, {del Burgo}, {Castro-Rodriguez}, {Corsini}, {D'Onghia},
  {Kundert}, {M{\'e}ndez-Abreu}, \& {S{\'a}nchez-Janssen}}]{zarattini2016}
{Zarattini}, S., {Girardi}, M., {Aguerri}, J.~A.~L., {et~al.} 2016, \aap, 586,
  A63

\bibitem[{{Zheng} {et~al.}(2007){Zheng}, {Bell}, {Papovich}, {Wolf},
  {Meisenheimer}, {Rix}, {Rieke}, \& {Somerville}}]{2007Zheng}
{Zheng}, X.~Z., {Bell}, E.~F., {Papovich}, C., {et~al.} 2007, \apjl, 661, L41

\bibitem[{{Zibetti} {et~al.}(2009){Zibetti}, {Pierini}, \&
  {Pratt}}]{zibetti2009}
{Zibetti}, S., {Pierini}, D., \& {Pratt}, G.~W. 2009, \mnras, 392, 525

\end{thebibliography}



\end{document}